\documentclass[twocolumn,pra,aps,showpacs]{revtex4}

\usepackage{amsmath}
\usepackage{amssymb}
\usepackage[dvips]{graphicx}
\usepackage{psfrag}

 \newtheorem{thm}{Theorem}[section]
 \newtheorem{lem}[thm]{Lemma}
 \newtheorem{cor}[thm]{Corolary}
 \newtheorem{defn}[thm]{Definition}
 \newtheorem{prop}[thm]{Proposition}
 \newenvironment{proof}{\noindent \emph{Proof.}}{\hspace*{\fill}$\square$}

 \newcommand{\n}{\mathbf{N}}
 \newcommand{\z}{\mathbf{Z}}
\newcommand{\Mat}{\mathbf{M}}
 \newcommand{\zd}{\z_D}
 \newcommand{\zdn}{\zd^n}
 \newcommand{\zddn}{\zd^{2n}}
 \newcommand{\zddm}{\zd^{2m}}

 \newcommand{\vect}[1]{\boldsymbol{\mathrm{#1}}}
\newcommand{\set}[2]{ \{\,#1:  #2\,\}}
 \newcommand{\sset}[1]{ \{#1\} }
 \newcommand{\half}{\frac 1 2}
 \newcommand{\transp}{^t}
 \newcommand{\lin}{\mathrm{Lin}\,}
 \newcommand{\rango}{\mathrm{rank}}
 \newcommand{\traza}{\mathrm{Tr}}

 \newcommand{\suma}[1]{{  {  \underset {\text{\raisebox{-.2ex}{$#1$}}}
  {\raisebox{-.6ex}{\text{\LARGE $\mathcal S$}}}  }  }}
 \newcommand{\sumatext}[1]{{\text{\Large $\mathcal S$}}_{#1}\,}
 \newcommand{\fase}[1]{\varphi(#1)}
 \newcommand{\twirl}[1]{T_{#1}}

 \newcommand{\ket}[1]{|#1\rangle}
 \newcommand{\bra}[1]{\langle #1|}
 \newcommand{\ketbradif}[2]{\ket{#1}\bra{#2}}
 \newcommand{\ketbra}[1]{\ketbradif {#1}{#1}}
 \newcommand{\biket}[2]{\ket{#1\;#2}}
 \newcommand{\bibra}[2]{\bra{#1\;#2}}
 
 \newcommand{\signoBell}{{\scriptscriptstyle{\mathcal B}}}
 \newcommand{\marcabell}{\text{\raisebox{-.5ex}{$\signoBell$}}}
 \newcommand{\preespacio}{\!\text{\hspace{0.2 ex}}}
 \newcommand{\postespacio}{\!\text{\hspace{0.25 ex}}}
 \newcommand{\postmarcabell}{\preespacio\marcabell}
 \newcommand{\premarcabell}{\marcabell\postespacio}
 \newcommand{\intermarcabell}{\preespacio\marcabell\postespacio}
 \newcommand{\sbell}[1]{\ket{#1}\postmarcabell}
 \newcommand{\sbellbra}[1]{\premarcabell\bra{#1}}
 \newcommand{\sbellketbradif}[2]{\ket{#1} \intermarcabell \bra {#2}}
 \newcommand{\sbellketbra}[1]{\sbellketbradif{#1}{#1}}
 \renewcommand{\bell}[2]{\biket{#1}{#2}\postmarcabell}
 \newcommand{\bellbra}[2]{\premarcabell \bibra{#1}{#2}}
 \newcommand{\bellketbra}[2]{\biket{#1}{#2} \intermarcabell \bibra {#1}{#2}}
\newcommand{\bellP}[2]{\biket{#1}{#2}\preespacio\text{\raisebox{-.5ex}{$\scriptscriptstyle{\mathcal B'}$}}}

 \newcommand{\Pmodelo}[1]{\mathcal{P}_{\scriptscriptstyle{\mathrm{#1}}}}
 
 \newcommand{\Pl}{\Pmodelo {loc}}
 \newcommand{\Ps}{\Pmodelo S}
 \newcommand{\Pt}{\Pmodelo T}
 \newcommand{\pt}[1]{\pi_{\vect {#1}}}
 \newcommand{\ps}[1]{\pi_{\scriptscriptstyle {#1}}}
 \newcommand{\psum}[1]{\pi_+^{#1}}
 \newcommand{\pbsum}[2]{\pi_+^{#1#2}}
 \newcommand{\pex}[1]{\pi_{\mathrm{xch}}^{#1}}
 \newcommand{\pswap}[2]{\pi_{\mathrm{swap}}^{#1#2}}
 \newcommand{\pdsum}[2]{\pi_{++}^{#1#2}}
 \newcommand{\ua}{U_\mathrm{A}}
 \newcommand{\ub}{U_\mathrm{B}}
 \newcommand{\Umodelo}[1]{\mathcal{U}_{\scriptscriptstyle{\mathrm{#1}}}}
 \newcommand{\Ul}{\Umodelo {loc}}
 \newcommand{\Ubl}{\Umodelo {B\,loc}}
 \newcommand{\Ubi}{\Umodelo {B\,inv}}
 \newcommand{\ugen}[1]{u_{\vect #1}}
 \newcommand{\ubi}[1]{U_{\vect #1}}
 \newcommand{\U}{\mathcal U}

 \newcommand{\h}{\mathcal{H}}
 
 \newcommand{\M}{M}
 \newcommand{\sym}{\,\Omega\,}
 \newcommand{\prodSym}{\transp\sym}
 \newcommand{\indep}[2]{\phi_{#1}(#2)}
 \newcommand{\divis}[1]{\mathrm{div}(#1)}
 \newcommand{\partic}[1]{C_{#1}}
 \newcommand{\particion}[3]{\partic{#1}(#2,#3)}
 \newcommand{\mcd}{\gcd}
 \newcommand{\perm}{\pi}
 \newcommand{\permas}[2]{\mu(#1,#2)}
 \newcommand{\permbs}[2]{\nu(#1,#2)}
 \newcommand{\perma}[2]{\permas{\vect #1}{\vect #2}}
 \newcommand{\permb}[2]{\permbs{\vect #1}{\vect #2}}
 \newcommand{\fasePerms}[2]{\phi(#1,#2)}
 \newcommand{\barfasePerms}[2]{\tilde{\phi}(#1,#2)}
 \newcommand{\fasePerm}[2]{\fasePerms{\vect {#1}} {\vect {#2}}}
 \newcommand{\oper}[1]{\mathcal{O}_{#1}}
 \newcommand{\mapP}{U}
 \newcommand{\dtrop}[1]{\rho_{\scriptscriptstyle #1}}
 \newcommand{\multirho}[1]{\rho^{\scriptscriptstyle(#1)}}
 \newcommand{\multip}[1]{p^{\scriptscriptstyle(#1)}}
 \newcommand{\multif}[1]{F^{\scriptscriptstyle(#1)}}
 \newcommand{\prob}{P}
 \newcommand{\VM}{V_{\scriptscriptstyle {\M}}}
 \newcommand{\pol}{\chi}

\begin{document}

\title[Short Title]{
Entanglement Distillation Protocols and Number Theory}

\author{H. Bombin and M.A. Martin-Delgado}
\affiliation{
Departamento de F\'{\i}sica Te\'orica I, Universidad Complutense,
28040. Madrid, Spain.
}

\begin{abstract}
We show that the analysis of entanglement distillation protocols for
qudits of arbitrary dimension $D$ benefits from applying
basic concepts from number theory, since the set $\zdn$ associated to Bell
diagonal states is a module  rather than a vector space.
We find that a partition of $\zdn$ into divisor classes  
characterizes the invariant properties of mixed Bell
diagonal states under local permutations. 
We construct a very general class of recursion protocols 
by means of unitary operations implementing
these local permutations.
We study these distillation protocols depending
on whether we use twirling operations in the intermediate steps or not, 
and we study them both analitically and numerically with Monte Carlo methods.
In the absence of twirling operations,
we construct extensions of the quantum privacy algorithms 
valid for secure communications with qudits of any dimension $D$.
When $D$ is a prime number, we show that  distillation protocols are
optimal both qualitatively and quantitatively.
\end{abstract}

\maketitle

\section{Introduction}
\label{sec1:level1}

Quantum Information Theory (QIT) revolves around the concept of
entanglement \cite{review1}, \cite{review2}, \cite{book}, \cite{rmp}. 
It is the product of combining the superposition principle
of Quantum Mechanics with multipartite systems -- described by tensor product
of Hilbert spaces. Entanglement is central to transmiting information in
a quantum communication protocol or processing information in a quantum
computation. There are two basic open problems in the study of entanglement:
separability and distillability. Separability is concerned with two
questions, namely, whether a quantum state is factorizable and if not,
how much entanglement it contains. These questions are of great importance
even in practice since entanglement amounts to interaction between two or more
parties, and thus it demands more resources to establish an 
entangled state than a factorized one.

Likewise, distillability \cite{bennett1}, \cite{super}, \cite{qpa}, 
\cite{chiara1}, \cite{horodeckis}, \cite{permutations}, \cite{beyond}
is concerned with two questions: whether a 
quantum state is distillable, and if it is, how to devise an explicit
protocol to extract entanglement out of the initial low entangled state.
The main focus of our paper is on the construction of distillation protocols,
rather than a direct analysis of the distillability issue.

The study of distillation protocols for qudits is justified since it is
known that for mixed states of dimension higher than $2\times 2, 2\times 3$, 
it is neither known a complete critierion for separability nor for distillability 
\cite{horodeckis}.

Separability and distillability are interconnected. Entanglement of a
mixed state is a necessary condition for being distillable. However, it was
quite a surprise to find that there exist states that though entangled,
they cannot be distilled. They are the bound entangled states that are
characterized by being PPT (Positive Partial Transposition)  entangled
states \cite{pavel}, \cite{bound1}, \cite{bound2}. 
This situation soon raised the question of whether Non-PPT states
were all distillable. Although there is not a conclusive answer, yet there
is strong evidence that this is neither the case since Werner states which
are finite-$n$ undistillable have been found \cite{evidence1}, \cite{evidence2}.

In this paper we study entanglement distillation protocols for qudits 
using the recursion method \cite{bennett1}, \cite{super}. The mixed states
to be distilled are diagonal Bell states of qudits, but they do not need to be
tensor product of pairs of Bell states. Moreover, we can also distill non-diagonal 
states.

We make significative progress in the understanding of these protocols 
and find new efficient
variants of them by using number theory. This number theory enters in the properties
of the module $\zdn$ that appears in the labelling of the Bell diagonal
sates of qudits. Local permutations acting on these states by means of unitary
operations serve to construct generalized distillation protocols 
\cite{permutations}.
As a byproduct, we also introduce heterotropic states \eqref{twirlPsResultado}
as the invariant states under the group of local permutations.

As a result of this study, we find that qudit states with dimension $D$ a prime
number are qualitatively and quantitatively the best choices for quantum distillation
protocols based on the recursion method. Qualitatively, since we show that for
$D$ not a  prime number there appear disturbing attractor points in the space
of fidelity parameters that deviate the distillation process from the desired
fixed point that represents the maximally entangled state.
This phenomenon is absent when $D$ is a prime number and $\zdn$ becomes a vector space.
Quantitatively, since we propose distillation protocols 
that when $D$ is a prime number, they distill all states with fidelity bigger than
$\frac{1}{D}$ without resorting to twirling operations.

We hereby summarize briefly some of our main results:

\noindent i/ We proof that the group of local permutations for qudits
of arbitrary dimension $D$ is the semidirect product of the group of
translations and simplectic transformations. This structure plays an 
important role in the distillation protocols for qudits.

\noindent ii/ We simplify  the problem of finding the best
distillation protocol to that of finding the best set of coefficients
of a certain polinomial constrained to the existence of a suitable vector space.

\noindent iii/ We introduce the concept of {\em joint performance parameter}
$\eta$ \eqref{eta} that allows us the comparison of distillation protocols 
with different
values of fidelity, probability of success and number of Bell pairs used altogether.
It is a figure of merit for low fidelity states above the distillation threshold
where the recursion method is specially suited for distillation, prior to 
switching a hashing or breeding method.

\noindent iv/ We analyze several distillation protocols assisted with twirling
operations as the dimension $D$ of qudits vary. We find that 
the best performance according to $\eta$ is not achieved for qubits ($D=2$), but for
qutrits ($D=3$) and $n=3$ input pairs of Bell states as shown in 
Fig.~\ref{GraphEta} and Fig.~\ref{GraphD3}.
We  also find that it is not possible to  improve  $\eta$ by indefinitely
increasing the number of input pairs $n$.

\noindent v/ We propose a distillation protocol without resorting to twirling 
operations for $n=4$ input Bell states and $m=2$ output Bell states that is
iterative and its yield is greatly improved:  about four orders of magnitude
with respect to the protocols based on twirling, 
even for states quite near the fixed point.
This is shown in Fig.~\ref{GraphD3Deutsch}.

\noindent vi/ We propose and study an extension of the Quantum Privacy Amplification
protocols \cite{qpa} that work for arbitrary dimension $D$. 

\noindent vii/ We find indications of the existence of non-distillable NPPT states
by studying the distillation capacities of protocols for several values of $D$,
as shown in Fig.~\ref{GraphOptimDist}.

This paper is organized as follows: In order to facilitate the reading and
exposition of our results, we present the proofs of our theorems and 
technicalities of the numerical methods in independent appendices.
Sect.~\ref{sec2:level1} deals with the basic properties of diagonal Bell 
states for qudits and introduce a partition of the module $\zdn$.
Sect.~\ref{sec3:level1} treats the group of local permutations acting
on qudits in diagonal Bell states.
Sect~\ref{sec4:level1} describes the group of local permutations and 
the twirling operations associated with it.
We characterize states that are invariant
under these operations as heterotropic states.
In Sect.~\ref{sec5:level1} we present all our distillation
protocols based on local permutations of qudits in diagonal Bell states, both
with and without resorting to twirling operations. To this end, we make extensive
use of the theoretical results found in previous sections and devise numerical
methods to analyse efficiently the properties of 
the proposed distillation protocols
as different parameters such as $D$, $F$, $n,m$ etc. vary.
Sect.~\ref{sec6:level1} is devoted to conclusions and future prospects.

\section{Basic Properties of Diagonal Bell States for Qudits}
\label{sec2:level1}

\subsection{Bell States Basic Notation}
\label{sec2:level2}

The quantum systems we are going to consider are \emph{qudits},
which are described by a Hilbert space of dimension $D \geq 2$,
and finite. The elements of a given orthogonal basis can be
denoted $\ket x$ with $x=0,\dots,D-1$. This set of numbers is
naturally identified with the elements of the set modulus
\begin{equation}
\zd := \z/D\z,
\end{equation}
and we shall informally use them as if they belonged to $\zd$. In
general, whenever an element of $\zd$ appears in an expression,
any integer in that expression must be understood to be mapped to
$\zd$.

We consider two separated parties, Alice and Bob, each of them
owning one of these systems. The entire Hilbert space then is $\h
= \h_A \otimes \h_B$. A mixed state of the whole system is called
\emph{separable} when it can be expressed as a convex sum of
\emph{product states} \cite{werner}:
 \begin{equation}\label{separable}
  \rho = \sum_{i} p_i \ketbra {e_i} \otimes \ketbra{f_i}, \ \  \ket {e_i}\in
  \h_A,\, \ket {f_i} \in \h_B;
 \end{equation}
A state which is not separable is said to be \emph{entangled}.

Elements of the \emph{computational basis} of a pair of qudits
shared by Alice and Bob are denoted as
\begin{equation}
 \biket{i}{j} := \ket{i}\otimes\ket{j}, \qquad i,j\in\zd.
\end{equation}
To shorthen the notation, it is convenient to introduce the symbols
\begin{align}
 \suma{k} := \frac{1}{\sqrt{D}}{\sum_{k\in\zd}}, \quad
 \delta(k) := \sqrt{D}\delta_{k,0}, \quad
 \fase{k} := {\rm e}^{\frac{2\pi {\rm i}}{D}k},
\end{align}
chosen so that $\sumatext{k}\fase{ik} = \delta(i)$ and
$\sumatext{k}\delta(i-k)f(k) = f(i)$ $(i\in\zd)$. \emph{Bell
states} are defined as 
\cite{gisin1}, \cite{gisin2}, \cite{mamd1}, \cite{mamd2}
\begin{equation}\label{Bell}
 \bell{i}{j} := \suma{k} \fase{ki} \biket{k}{k-j} \qquad i,j\in\zd.
 \end{equation}
Bell states are an example of maximally entangled states. In fact,
any maximally entangled state can be identified with the $\bell 0
0$ state by suitably choosing the computational basis of each of the
qudits, due to the Schmidt decomposition.

The \emph{fidelity} of a mixed state $\rho$ is defined as
\begin{equation}
F:=\max_{\Psi} \bra {\Psi} \rho \ket {\Psi}, 
\label{fidelity}
\end{equation}
where the maximum is
taken over the set of maximally entangled states. The aim of
distillation protocols is to get maximally entangled pairs
(fidelity one) by means of local operations and classical
communication (LOCC), starting with entangled states of fidelity
lower than one. Because of the previous comment, we will always
suppose, without lose of generality, that the initial fidelity of the
states to be distilled is equal to $\bellbra 0 0 \rho \bell 0
0$, and the aim of our protocols will be to obtain distilled states
as close as possible to this Bell state.

 Of special interest are the mixtures of perfectly entangled states and white
 noise, known as \emph{isotropic states}:
\begin{equation}\label{isotropo}
\rho_{\rm iso}:= F \: \bellketbra 0 0 + \frac {1-F}{D^2-1}\:
  \bigl (
   1 - \bellketbra 0 0
  \bigr ),
\end{equation}
 where $F$ is the fidelity
of the state. These states are known to be entangled and
distillable iff $F>\frac 1 D$ \cite{horodeckis}.

The main interest of these states comes not only from their
physical meaning, but also from the possibility of transforming any
state in an isotropic one through a \emph{twirling} operation. In
general, the  twirling consists in a random application of the elements
of a certain group of unitary operations, say $\U$, to each of the
systems in an ensemble. Namely, its action is
\begin{equation}\label{twirl}
 \twirl{\U} (\rho) := \int_{\U} dU \, U \rho \,U^\dagger.
\end{equation}
The result of such an operator must be a sum over the states
invariant under the action of the group. In the case of isotropic
states, a suitable election is the set of transformations of the
form $U\otimes U^\ast$.

When managing multiple shared pairs, vector notation is necessary;
$\vect{k}\in\zdn$ stands for $\vect{k} = (k_1, \dots , k_n),
k_i\in\zd$. Scalar product will be employed with its usual
meaning. The generalization of the previous expressions is
straightforward:
\begin{equation}
 \suma{\vect{k}} := \suma{k_1} \dotsi \suma{k_n}, \quad
 \delta(\vect{k}) := \delta(k_1) \dotsi \delta(k_n).
\end{equation}
Again, $\sumatext{\vect{k}}\fase{\vect{i}\cdot\vect{k}} =
\delta(\vect{i})$ for any $\vect{i}\in\zdn$. The computational
basis and the Bell basis are:
\begin{align}\label{nBell}
 \biket{\vect{i}}{\vect{j}} &:= \overset{n}{\bigotimes_{k=1}}\,
 \biket{i_k}{j_k},\notag \\
 \bell{\vect{i}}{\vect{j}} &:= \overset{n}{\bigotimes_{k=1}}\, \bell{i_k}{j_k} = \suma{\vect{k}} \fase{\vect{i}\cdot\vect{k}}
 \biket{\vect{k}}{\vect{k}-\vect{j}},
\end{align}
with $\vect{i},\vect{j}\in\zdn$. In order to simplify the
notation, sometimes we will work with vectors over $\z_D^{2n}$ and
write states
as $\ket {\vect x}$ in the place of $\biket {\vect i}{\vect j}$, with
\begin{equation}\label{CorrespondenciaZddnZdn}
 \vect x := (i_1,\dots, i_n, j_1,\dots,j_n).
\end{equation}

\subsection{A Partition of  $\zdn$ with Divisor Classes}
\label{sec2:level3}

In general, $\z_D$ is not a field and thus $\z_D^n$ is not a vector
space, but a module. We can still make use of some properties associated to
vector spaces and so we will abuse a bit of the term vector. For a detailed
exposition see appendix \ref{ApendiceVector}, but it is enough to
know the following. The usual definition of linear independence
makes sense, as one can demonstrate that any linearly independent
set of vectors can be extended to a complete base and also that a
square matrix composed by such a set is invertible. A subspace is
defined to be the set of linear combinations of a linearly
independent set, and its dimension is the cardinality of these set
of generators. Orthogonality poses no problem, since the set
orthogonal to a subspace is a subspace, and it has the expected
dimension.

Working with this pseudo-vector space $\zdn$ requires care. Some vectors
can be taken to the null vector by multiplying them by a non-zero
number. For example, for $D=4$ we have $2\cdot(0,2) = (0,0)$. In
order to classify this anomalous vectors, consider the set of
divisors of D:
\begin{equation}
 \divis D := \set {d\in\n}{d|D}.
\end{equation}
This set inherits the ordering of $\n$, and we shall use this
property to introduce a suitable $\mcd$ function in $\zd$:
\begin{defn}\label{defmcd}
For every $S\subset \zd$ we define the greatest common divisor of
$S$, or $\mcd(S)$, to be the greatest $d \in \divis D$ such that
$\frac D d s = 0$, $\forall\,s\in S$.
\end{defn}
The nomenclature was chosen because for any $d\in\divis D$ and
$x\in \z$ we have
\begin{equation}
 d|x \iff D|\frac D d x \iff \frac D d x = 0\pmod D,
 \end{equation}
and then for any set of integers $X$
\begin{equation}\label{defmcdAlternativa}
 \mcd(\overline X) = \max \set {d\in\divis D}{d|x\:\forall\,x\in
 X},
\end{equation}
where $\overline X$ is the corresponding set in $\zd$.

\begin{table}
\begin{ruledtabular}
\begin{tabular}{|c||c|c|c|c|c|}
   $D$ & $2$ & 3 & 4 & 5 & 6 \\
  \hline
  \hline
  $\phi_1(D)$ &1 & 2 & 2 & 4 & 2 \\  \hline
 $\phi_2(D)$  &3 & 8 & 12 & 24 & 24 \\ \hline
 $\phi_3(D)$  &7 & 26 & 56 & 124 & 182 \\ 
\end{tabular}
\end{ruledtabular}
\caption{Values of the generalized Euler's totient function $\indep n D$
for several qudit dimensions $D$.}
\label{tablaPhi}
\end{table}

\begin{table*}
\begin{ruledtabular}
\begin{tabular}{|c|c|c|c|c|c|}
   D & 2 & 3 & 4 & 5 & 6 \\
  \hline
  \hline
  $\sharp\Ps(D,1)$ &6 & 24 & 48 & 120 & 144 \\ \hline
  $\sharp\Ps(D,2)$ & 720 & 51,840 & 737,280& $9.36 \cdot 10^6$ & $\sim3.7\cdot 10^7$ \\ \hline
   $\sharp\Ps(D,3)$ &1,451,520 & $\sim9.2\cdot 10^9$ & $\sim3.0\cdot 10^{12}$  & $\sim9.1\cdot 10^{13}$ & $\sim1.3\cdot 10^{16}$ \\
\end{tabular}
\end{ruledtabular}
\caption{Values of the number of elements of the group $\Ps(D,n)$
for several qudit dimensions $D$.}
\label{tablaPs}
\end{table*}

Vectors over $\zd$ are n-tuples of elements in $\zd$, and so we
extend the $\mcd$ function to act over $\zdn$ in the natural way,
that is, if $\vect v = (v_1,\dots\,v_n)$, $\mcd(\vect v) :=
\mcd(\sset{v_1,\dots, v_n})$. Now we can consider an equivalence
relation in $\zdn$ governed by the equality under the $\mcd$
function. The corresponding partition consists in the sets:
\begin{equation}\label{particion}
\particion d D n := \set {\vect v \in \zd^{n}}{\mcd(\vect v) = d}, \ d\in
\divis D
\end{equation}
The most important of these sets is $\particion 1 D n$, since it
contains those vectors $\vect v$ for with $\sset {\vect v}$ is
linearly independent. Later we will need its cardinality when considering
properties of local unitary operators acting on diagonal Bell states. Thus,
it is useful to define:
\begin{equation}\label{defPhi}
 \indep n D :=
 \begin{cases}
  1 & \text{if $D=1$} \\
  \sharp \particion 1 D n & \text{if $D>1$}
 \end{cases}
\end{equation}
For the particular case of $n=1$,
$\indep 1 x$ corresponds to  Euler's totient $\phi$-function \cite{Euler}.
Euler's $\phi$-function appears naturally in number theory since
it gives for a natural number $n$, the
cardinality of the set $\set {m=1,\dots,n-1}{\gcd(m,n)=1}$.
That is,  $\indep 1 n$ is the total number of 
coprime
integers (or totatives) 
below or equal to $n$. 
For example, there are eight totatives of 24, namely, 
$\sset{1, 5, 7, 11, 13, 17, 19, 23}$, thus $\phi_1(24) = 8$.
For $n\neq 1$, we have therefore introduced a generalization of Euler's totient function
for elements in $\zdn$. The following lemma gives us how to compute the cardinalities
of the sets $\particion d D n$, 
which shall naturally arise in our analysis of distillation protocols. 
\begin{lem}\label{lemaPhi}
    For every $n\in\n$, $D\in \n-\sset 1$ and $d\in\divis D$:
 \begin{enumerate}
    \item
    \begin{equation}\label{valorphi}
     \indep n D = D^n \prod_{p|D \atop p\:\mathrm{prime}} \frac {p^n-1}{p^n}
    \end{equation}
    \item
    \begin{equation}\label{cardinalParticion}
     \sharp \particion d D n = \indep n {\frac D d}
    \end{equation}
    \item
    \begin{equation}
     \sum_{d'\in\divis D} \indep n {d'} = D^n
    \end{equation}
\end{enumerate}
\end{lem}
The proof of this lemma can be found in
appendix \ref{ApendicePhi}. As an illustration,
we list several values of $\indep n D$ in Table~\ref{tablaPhi}.

\section{The group of local permutations}
\label{sec3:level1}

The main constraint Alice and Bob have to face when they intend to
distill qudits is that they can perform only local operations. If
we consider only unitary operations, we are led to the group $\Ul$
of local unitary operations. Its elements are all of the form
\begin{equation}
U=\ua\otimes\ub.
\end{equation}
In this section we shall study the subgroup $\Ubl$, defined as the
group of local unitary operations which are closed over the space
of Bell diagonal states, that is, states of the form
\begin{equation}\label{BellDiagonal}
 \multirho n= \sum_{\vect x\in\zddn} \multip n_{\vect x} \,\sbellketbra {\vect
 x},
\end{equation}
where the label $(n)$ is a reminder that we are considering states
of $n$ pairs of qudits. The aim is to use the acquired knowledge
to devise distillation protocols specially suited for these
states.

More specifically, we analyze the group $\Ubl(D,n)$ of local
unitary operators over the space spanned by the tensor product of
$n$ pairs of qudits of dimension $D$ for which the image of a Bell
diagonal state is another Bell diagonal state. The first we notice
is that the result of applying such an operator over a pure Bell
state is another Bell state (it cannot be the convex sum of
several Bell states because it must remain pure). Since the
mapping of Bell states must be one to one, the action of any
$U\in\Ubl(D,n)$ involves a permutation of the Bell states:
\begin{equation}\label{BellDiagonalTransformado}
 U \rho \,U^\dagger = \sum_{\vect x\in\zddn} \multip n_{\vect x} \,\sbellketbra {\perm(\vect
 x)},
\end{equation}
where $\perm:\zddn\rightarrow\zddn$ is a permutation. So we
introduce $\Pl(D,n)$, the \emph{group of local permutations}, as
the set of permutations over $\zddn$ implementable over Bell
states by local (unitary) means.

 Before stating the main result of this section, we shall
 define several groups.
Consider the family of unitary operators $\ugen x$ ($\vect x \in
\zddn$) over Bob's part of the system such that by definition
\begin{equation}\label{defGeneraBell}
1 \otimes \ugen x^\ast \,\sbell {\vect 0} := \sbell {\vect x},
\end{equation}
where conjugation is taken respect to the computational basis.
With this operators at hand, we construct the group $\Ubi(D,n)$
with the elements $\ubi x := \ugen x \otimes {\ugen x}^\ast$. We
claim that it is a subgroup of $\Ubl(D,n)$. An explicit
calculation shows that the action of its elements is
\begin{equation}\label{accionUbi}
 \ubi x \,\sbell {\vect y} =
 \fase{\vect x \prodSym \vect y} \,\sbell {\vect y},
\end{equation}
where $\Omega\in\Mat_{2n\times 2n}(\zd)$ is
\begin{equation}\label{defOmega}
 \sym:=\begin{bmatrix} 0 & 1 \\ -1 & 0 \end{bmatrix}.
\end{equation}
So the special feature of $\Ubi$ is that for $\multirho n$ Bell
diagonal $\ubi x \multirho n\ubi x^\dagger = \multirho n$, which
means that its elements implement the identity permutation.

 We also define two subgroups of the group of permutations over
 $\zddn$. The \emph{translation group} $\Pt(D,n)$ contains the
 permutations of the form
 \begin{equation}\label{Pt}
 \pt a(\vect x) = \vect x + \vect a,
 \end{equation}
 with $\vect a\in\zddn$, and the \emph{symplectic group} $\Ps(D,n)$ contains in
 turn those whose action is
\begin{equation}\label{Ps}
 \ps \M(\vect x) = \M\vect x,
 \end{equation}
 where $\M\in \Mat_{2n\times 2n}(\zd)$ is such that
\begin{equation}\label{CondicionGrupoM}
 \M\transp\sym \M=\sym.
 \end{equation}

$\Ps(D,n)$ is a   finite non-simple group. A suitable generator set for this
group is presented in appedix \ref{ApendiceGeneraPs}.
Now, we are in position to establish the following theorem that plays
an important role in the distillation protocols for qudits to be 
devised later on.

\begin{thm}\label{thmPlocal}
$ $

\begin{enumerate}
 \item
  $\Pl$ is the semidirect product of $\Ps$ and $\Pt$:
 \begin{equation}
  \Pl(D,n) = \Pt(D,n)\ltimes\Ps(D,n)
 \end{equation}
 \item
  Given $U_1,U_2\in\Ubl(D,n)$ such that they yield the same permutation
  in $\Pl$, there exists $\ubi x\in\Ubi(D,n)$ such that
 \begin{equation}\label{globalphase}
  U_1 U_2^{-1} = \phi\, \ubi x
 \end{equation}
 where $\phi$ is a phase.
\end{enumerate}
\end{thm}

We prove this theorem in appendix \ref{ApendiceProofThm}.
For qubits ($D=2$), part 1 of this theorem was proved in \cite{permutations}
using a mapping between Bell states and Pauli matrices. Our proof does
not rely on this mapping and being completely different,  it becomes
general enough so as to treat all qudits of dimension $D$ on equal footing.

$\Ps$ is specially well suited to construct distillation protocols, and
so it is worth a closer study of its properties. There is another
interesting way of writing \eqref{CondicionGrupoM}; if we call
$\vect u_i$ the first $n$ rows (columns) of $M$ and $\vect v_i$
the last $n$ rows (columns), the condition can be rewritten in a
canonical simplectic form:
\begin{align}\label{Canonicas}
 \vect u_i \prodSym \vect u_j &= 0 \notag\\
 \vect v_i \prodSym \vect v_j &= 0 \\
 \vect u_i \prodSym \vect v_j &= \delta_{ij}\notag
\end{align}
This point of view is specially useful when systematically
constructing the elements of $\Ps$, thanks to the following result:
\begin{thm}\label{thmCardinalPs}
$ $

\begin{enumerate}
 \item
 Consider a linearly independent set of vectors $\sset{\vect u_1,
 \dots,\vect u_r, \vect v_1, \dots,\vect v_s, \vect v_{r+1}, 
 \dots, \vect v_{r+t}}\subset\zddn$ with ($s\leq r\leq n,s+t\leq n$).
 If this set satisfies conditions \eqref{Canonicas}, it is always
 possible to complete it preserving them.
\item
  The cardinality of $\Ps$ is
  \begin{equation}
   \sharp\Ps(D,n) = D^{n^2}\prod_{k=1}^{n} \indep{2k}{D}
  \end{equation}
\end{enumerate}
\end{thm}

We prove this theorem in appendix \ref{ApendiceCardinalPs}.
As an illustration,
we list several values of $\sharp\Ps(D,n)$ in Table~\ref{tablaPs}. 
Clearly numbers grow
fast, which makes unfeasible  any numerical investigation
which requires going over the elements of the whole group even for not very
large values of $n$. In any case,
it can be helpful to have an algorith which allows this task without the expense of 
storing the elements. Consider any suitable ordering over $\zddn$. Given an element of
$\Ps(D,n)$, we can increase its last row according to this order till another element
is reached. If this fails, the same can be done for the previous row, and so on and so forth. However, 
this is not very efficient, and we can do it better combining part one of theorem \ref{thmCardinalPs}
with lemma \ref{lemaGrupo}. For example, for any of the last $n$ rows we could substitute 
the search with the application of a suitable generator from lemma \ref{lemaGrupo}. Then the additional
information contained in the proof of theorem \ref{thmCardinalPs} would guarantee that we were not forgetting
any element of the group. Moreover, as we shall later see, tipically we are only interested in some of the
rows of the matrix, and then part 1 of the theorem is crucial since it allows usto ignore unimportant rows.

\section{Twirling with $\Ubi$ and $\Pl$}
\label{sec4:level1}

We now explore the possibility of using the groups of the
previous section with the twirling operator \eqref{twirl}, which
for finite groups is:
\begin{equation}\label{twirlFinito}
 \twirl{\U} (\rho) := \frac 1 {\sharp \U} \sum_{U\in\U} \, U \rho \,U^\dagger.
\end{equation}
Consider now the group $\Ubi(D,n)$. From \eqref{accionUbi}, it follows
\begin{equation}
 \suma {\vect z} \,\ubi z \,\sbellketbradif {\vect
x} {\vect y} \,\ubi z^\dagger = \delta(\vect x - \vect y)
\,\sbellketbra {\vect x},
\end{equation}
which means that the action of $\twirl{\Ubi}$ over a state leaves
Bell diagonal elements invariant, whereas the off-diagonal components are
sent to zero.

The group $\Ps$ can also be successfully used in twirling
operations. This asseveration, however, has no meaning by itself
since $\Ps$ is not a group of transformations over the $n$ pairs
of qudits. We have to choose any mapping $\mapP:\Ps\longrightarrow
\Ubl$ such that
\begin{equation}
 \mapP(\perm) \sbellketbra {\vect x} \mapP^{\dagger}(\perm)= \sbellketbra {\perm(\vect
 x)}.
\end{equation}
There are many possible realizations for this mapping,
and at least in general the image of the mapping is not a subgroup
of $\Ubl$.  However, it \emph{is} a group when considered as a set of
transformations over Bell diagonal states. Thus, for $\rho$ Bell
diagonal the following makes sense:
\begin{equation}\label{twirlPs}
 \twirl{\Ps} (\rho) := \frac 1 {\sharp \Ps} \sum_{\perm\in\Ps} \, \mapP(\perm) \rho \,\mapP(\perm)^\dagger.
\end{equation}
 To perform the sum we need to know which are the states invariant under the
action of the group.
\begin{thm}\label{thmMCDinvariante}
For every $\vect x, \vect y \in \zddn$, $\mcd(\vect x) =
\mcd(\vect y)$ if and only if there exists a permutation in
$\Ps(D,n)$ with associated matrix $M$ such that $\vect x = M \vect
y$.
\end{thm}
\begin{proof}
The if direction follows from $M$ being invertible, since then
$dM\vect{i}=Md\vect{i}=0$ iff $d\vect{i}=0$. We now proof the 
only if direction. Let $d = \mcd(\vect x)$ and consider any $\vect
u \in \zddn$ such that $d\vect u = \vect x$ (note that $\mcd(\vect
u)=1$).  $\sset{\vect u}$ is linearly independent, and so there
exists a matrix $M$ associated to a permutation in $\Ps(D,n)$
having $\vect u$ as its first column (see theorem \ref{thmCardinalPs}).
 Then, if $\vect v = (d, 0,\dots, 0)$ we
have $\vect x = M_1 \vect v$. The same reasoning is true for
$\vect y$, giving $\vect y = M_2\vect v$ and thus $M=M_1M_2^{-1}$.
\end{proof}

Now, let us recall the partition in $\zddn$ associated to the function
$\mcd$ (see \eqref{particion}). We define the related states
\begin{equation}\label{invariantes}
\dtrop d := \frac 1 {\sharp \partic d} \sum_{\vect x \in
\partic d} \sbellketbra {\vect x}.
\end{equation}
These are the invariant states we were searching for. Thus, if $\rho$
is Bell diagonal:
\begin{equation}\label{twirlPsResultado}
\twirl{\Ps} (\rho) = \sum_{d\in\divis D} \frac {\traza (\dtrop d
\rho)}{\traza (\dtrop d \dtrop d)} \dtrop d
\end{equation}
Note that we have not taken
into account the number of pairs involved, since it is unimportant.
However, usually twirling operations are interesting
for $n=1$. In this case, in analogy with isotropic states, we shall call 
heterotropic states
those states invariant under \eqref{twirlPsResultado}.
If $\rho$ is not Bell diagonal, we can still obtain the same
result with the operator
\begin{equation}\label{twirlUbiPs}
 \twirl{\Ubl\times\Ps} (\rho) := \frac 1 {\sharp \Ubi}\frac 1 {\sharp \Ps}
   \sum_{\perm\in\Ps}\sum_{U\in\Ubi} \, \mapP(\perm)U \rho \,U^\dagger\mapP(\perm)^\dagger.
\end{equation}
As a corollary, if $D$ is prime there are just two Bell diagonal invariant 
states
\begin{align}
\dtrop 1 &= \frac 1 {D^{2n}-1} \bigl ( 1-\sbellketbra{\vect 0} \bigr ), \\
\dtrop D &=\sbellketbra{\vect 0},
\end{align}
and thus the result of the twirling operation is an isotropic
state, 
which is the simplest example of an heterotropic state.

\section{Permutation assisted distillation}
\label{sec5:level1}

In the distillation protocols we consider, which are iterative,
each iteration cycle can be decomposed in the following steps:

\begin{em}
\begin{enumerate}
 \item 
At start, Alice and Bob share $n$ qudit pairs of dimension
       $D$ and state matrix $\multirho n$.
 \item They apply by local means one of the permutations $\ps \M\in\Ps(D,n)$
       in \eqref{Ps}.
 \item They measure the last $n-m$ qudit pairs, both of them in
       their computational basis.
 \item If the results of the measurement agree for each of the
       measured pairs, they keep the first $m$ pairs (in the state
       $\multirho m$). Else, they discard them.
\end{enumerate}
\end{em}

In most situations, the initial $n$ pairs are independent and have
equal state matrices $\rho$. In these cases
\begin{equation}\label{rhoEsProducto}
\multirho n = \rho^{\otimes n}.
\end{equation}
In general (for $m>1$) this does not guarantee that $\multirho m$
will be a product state, however, and thus it is preferable to
consider the most general case.

The process can be performed several times in order to improve the
entanglement progressively, but it is worth taking into account
that a scheme of this kind with $s$ steps and, say, $n=2$ and
$m=1$, is equivalent to a single-step one with $n'=2^s$ and
$m'=1$. It is enough to perform initially all the permutations and
afterwards all the measurements. Although convergence properties
are the same, the equivalence is not complete since from a
practical point of view the step by step method will give a better
yield. This is so because an undesired result in the measurement
is more harmful in the second case, as more pairs must be
discarded at once.

In appendix \ref{ApendiceDestilacion} we derive an expression for
the state of the remaining pairs of qudits after a successful
measurement. It appears that the protocol is blind to non diagonal
states (in the Bell basis). So let us define
\begin{alignat*}{2}{}
 \multip n_{\vect x} &:= \sbellbra {\vect x} \multirho n\sbell {\vect
 x}, \qquad &\vect x\in\zddn;\\
 \multip m_{\vect x} &:= \sbellbra {\vect x} \multirho m\sbell {\vect x},
  \qquad &\vect x\in\zddm.
\end{alignat*}

If we call $\VM$ the space generated by the last $n-m$ rows of
$\M$ (the matrix associated to $\ps {\M}$) the probability of
obtaining the desired measure is
\begin{equation}\label{probabilidadResultado}
 \prob = \sum_{\vect x \in \VM^\bot} \multip n_{\vect x},
\end{equation}
and the recurrence relation for the probabilities is
\begin{equation}\label{recurrenciaProbabilidades}
 \multip m_{\vect x} = \frac 1 {\prob}
  \sum_{\vect y \in \VM} \multip n_{\sym\vect y+\M^{-1}\vect {\bar x}},
\end{equation}
where $\vect x \in \zd^{2m}$ and $\vect {\bar x} \in \zd^{2n}$ is
\begin{equation}\label{defxBarrada}
\vect {\bar x} := (x_1,\dots,x_m,
\underset{n-m}{\underbrace{0,\dots,0}},x_{n+1},\dots,x_{n+m},\underset{n-m}{\underbrace{0,\dots,0}}).
\end{equation}
 Note that since $\M^{-1} = \sym\transp\M\transp\sym$, rows $m+1$ to $n$
 (of $\M$) do not take part in the expression, and therefore the protocol
does not depend on them.

Consider the following family of heterotropic states:
\begin{equation}
 \rho^{\mathrm{fix}}_d := \sum_{\vect x\in \zd^2} \frac{1}{Dd^2} 
 \delta (d \vect x) \sbellketbra{\vect x}, \qquad d\in\divis D.
\end{equation}
From theorem \ref{thmMCDinvariante} we know that
\begin{equation}
 \delta(d M\vect x) = \delta(d\vect x).
\end{equation}
Using this fact and equation \eqref{estadoFinalIntermedio}, one can readily
check that for $\multirho n = {\rho^{\mathrm{fix}}_d}^{\otimes n}$
\eqref{recurrenciaProbabilidades} gives
$\multirho m = {\rho^{\mathrm{fix}}_d}^{\otimes m}$. Therefore, these heterotropic states are
always fixed points of the protocol and candidates for attractors.

In the case of single-step protocols, we are only interested in the
final joint fidelity (the probability of the state being
$\sbellketbra {\vect 0}$)
\begin{equation}\label{fidelidadFinal}
 \multif m = \frac 1 {\prob}
 { \sum_{\vect x \in \VM} \multip n_{\sym\vect x} }
\end{equation}
where the label $(m)$ reminds us that this is the joint
probability of the $m$ pairs being in the desired state. In
general, the fidelity of each pair will be greater. If $m=1$ these
distinction vanishes, and we will simply write $F'$ instead of
$\multif 1$. Equations \eqref{fidelidadFinal} and
\eqref{probabilidadResultado} show how the effect of the entire
process relies only in the set $\VM$, thereby reducing the search for
efficient protocols according to part one of theorem
\ref{thmCardinalPs}.

We now consider the Fourier transform of the probabilities
\begin{equation}\label{transformadaProbabilidades}
 \multip n_{\vect {\tilde x}} : = \sum_{\vect x\in \zddn} \fase {\vect {\tilde x}\cdot \vect x } \multip n_{\vect x}, \qquad
  \vect {\tilde x} \in \zddn \\
 \end{equation}
to obtain
\begin{equation}\label{ProbabilidadSimple}
 P = D^{m-n} { \sum_{\vect {\tilde x} \in \VM} \multip n_{\vect {\tilde x}} },
\end{equation}
where we have used 
\begin{equation}\label{identidadSumaVectorSpace}
 \sum_{\vect v\in V} \fase{\vect v\cdot\vect u}:=
 \begin{cases}
  0 & \text{if $\vect u\not \in V^{\bot}$} \\
  \sharp V & \text{if $\vect u\in V^{\bot}$}
 \end{cases},
\end{equation}
with $V$ being  a subspace of $\zdn$ and $\vect u\in\zdn$.
Gathering these results, we have
\begin{equation}\label{fidelidadFinalSimple}
 \multif m = D^{n-m} \frac
 { \sum_{\vect x \in \VM} \multip n_{\sym\vect x} }
 { \sum_{\vect x \in \VM} \multip n_{\vect {\tilde x}} },
\end{equation}
an expression for the final (joint) fidelity which can lighten its
direct computation when \eqref{rhoEsProducto} holds, since for
this case we can define for $a,b\in\zd$:

\begin{align}
  p_{ab} := \bellbra ab \rho \bell ab,& \qquad
  p_{\tilde a\tilde b} := \sum_{a,b \in \zd} \fase {\tilde a a+\tilde bb}
  p_{ab}\label{defProbindiv}
\intertext{and then} \multip n_{\vect x} = \prod_{i=1}^n
p_{x_i,x_{n+i}},&
 \qquad \multip n_{\vect {\tilde x}} = \prod_{i=1}^n p_{\tilde x_i,\tilde x_{n+i}}
 \label{ProbabilidadProducto}
\end{align}

\subsection{Twirling assisted protocols}
\label{sectTwirling}

In order to understand better equation
\eqref{recurrenciaProbabilidades}, we will adopt a useful
simplification which generates by itself a whole family of
distillation protocols. We consider only initial states for which
the pairs are mutually independent and equal as in
\eqref{rhoEsProducto}. Moreover,  before each iteration we introduce any
twirling operation which leads to an isotropic state while
preserving the fidelity, as in \eqref{isotropo}. This way the
evolution of the state through the distillation protocol is
described entirely by a single parameter, the fidelity $F$.

We would like to evaluate \eqref{ProbabilidadProducto}. We need:
\begin{align}
  p_{ab} &= F \,\delta_{ab} + \frac {1-F}{D^2-1} \,(1-\delta_{ab}),\\
  p_{\tilde a\tilde b} &= \delta_{\tilde a\tilde b} + \frac{D^2 F - 1}{D^2-1}
  \,(1-\delta_{\tilde a\tilde b}).
\end{align}
Defining $c_1(F):=\frac {1-F}{F(D^2-1)}$ and $c_2(F):=\frac{D^2 F
- 1}{D^2-1}$, this implies that for any $\vect x$ we have $\multip
n_{\vect x}= F^{n}c_1(F)^s$ and $\multip n_{\tilde {\vect x}} =
c_2(F)^s$, where $n-s$ is the number of occurrences of $p_{00}$ in
the product (that is, $\sharp\set{r=1,\dots,n}{x_r=x_{n+r}=0}$).
 Since $\multip n_{\sym \vect x} = \multip n_{\vect x}$, we can write
\begin{align}
 \multif m &= D^{n-m} F^n\,\frac
 { \pol ( c_1(F) ) }
 { \pol ( c_2(F) ) }\label{recurrenciaFidelidadSimple}
 \intertext{and}
 P &= D^{m-n} \pol (c_2(F)).\label{ProbabilidadSimple2}
\end{align}
where $\pol(x) = \sum_{s=0}^n \lambda_s x^s$ is a polynomial with
coefficients defined by
\begin{equation}\label{coefsPolinomio}
 \lambda_s := \sharp \set {\vect x \in \VM} {n-s =
  \sharp \set {r=1,\dots,n} {x_r=x_{n+r}=0}}.
\end{equation}
This definition is not very useful when trying to construct a
suitable $\VM$ for a given $\pol$, but we can do better. Let
$V_r$ be the set of linear combinations of columns $r$ and $n+r$
of the matrix formed by the last $n-m$ rows of $M$, (or any other
matrix of size $(n-m)\times 2n$ such that its rows span $\VM$). If
$V_r$ is a subspace of $\zd^{n-m}$ for every $r$, which indeed 
is always the case for $D$ prime, we can rewrite
\eqref{coefsPolinomio} as
\begin{equation}\label{coefsPolinomioVectorial}
 \lambda_s = \sharp \set {\vect v \in \zd^{n-m}} {n-s = \sharp \set {r=1,\dots,n} {\vect v \in V_r^\bot} }
\end{equation}
We remark that $\pol$ depends only on $\VM$, which is a subspace
of $\zddn$ of dimension $n-m$ constrained only by
\begin{equation}\label{condicionVM}
\vect u \prodSym \vect v = 0\quad\forall \:\vect u,\vect v \in
\VM.
\end{equation}

It is apparent that $\pol(1) = D^{n-m}$ and $\pol(0) = 1$.
For $m=1$ \eqref{recurrenciaFidelidadSimple} becomes the recurrence relation
$F'=F'(F)$, and then among the fixed points are $F=1$ (perfect
entanglement), $F = \frac 1 D$ (maximum fidelity for separable
states) and $F = \frac 1 {D^2}$ (pure noise).

Therefore, we have reduced the problem of finding the best protocol to that
of finding the best coefficients for the polynomial, constrained
to the existence of a suitable vector space. In the next
subsection, we survey this issue for several values of $n$, but
previously  a small consideration is worthwhile. For the identity
permutation (which of course is completely useless) the
coefficients of the polynomial are $\lambda_s = {n-m \choose s}
(D-1)^s$ and therefore
\begin{equation}\label{probabilidadIdentidad}
 P = \left (\frac {1+(D-1)F} D\right)^{n-m}.
 \end{equation}
One expects the probability of a useful protocol to be less than
this, but then the decay is at least exponential respect to an
increase in $n-m$. This is an early advisory that considering
progressively larger values of $n$ needs not be better.

\begin{figure}
\psfrag{D}[Bc][Bc][1][0]{$D$}
\psfrag{eta}[Bc][Bc][1][0]{$\eta$}
\includegraphics[width=8.5 cm]{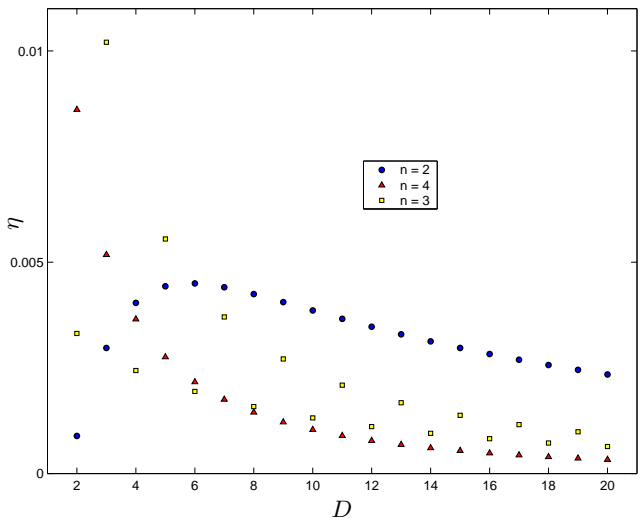}
\caption{
 Values of the coefficient $\eta$ for the considered protocols with $n=2,3,4$.
} \label{GraphEta}
\end{figure}

\subsubsection{Low fidelity states}

Now we will concentrate on low fidelity states near $F=\frac 1
D$. Since hashing and breeding protocols are available for high
fidelities \cite{super}, \cite{beyond}, one reason for studying this range is that it is the
natural testground for the class of protocols we are analyzing.
It is interesting also because we can develop a method to compare
in a simple manner protocols with different $n$.

We start by discarding protocols with $m>1$. The reason to do so
is that $F(\frac 1 D)=\frac 1 {D^m}$. Since one does not expect
the resulting pairs to be statistically uncorrelated, in general
the individual fidelities will be less than $\frac 1 D$, making
the algorithm useless near the point of interest. We will see
later how protocols with $m>1$ can be fruitfully used.

A problem arising when comparing different protocols is that
several factors take part at the same time, making it difficult to
balance them in a simple manner. In our case, we have to take into
account the probability of obtaining the right measure $P$, the
number of pairs used $n$ and the output fidelity $F'$. We will
now see, however, that restricting our attention to low fidelity
states allows us to introduce  a single coefficient which makes
possible the comparison. We shall call this coefficient the 
{\em joint performance} 
$\eta$ of a distillation protocol and it is
constructed as follows.

Let us consider an isotropic state of fidelity $\frac 1
D+\epsilon$. After $q$ steps of the protocol, at the lowest order
in $\epsilon$, the state will have a fidelity $\frac 1
D+F_1^q\epsilon$ and the yield will be $\bigl (\frac {P_0}
n\bigr)^q$, with
\begin{align}
F_1 &:= \frac {dF'}{dF}\Bigg\vert_{F=\frac 1 D}=n-\frac {2
D}{D^2-1} \left
[\frac d {dx} log(\pol(x))\right]_{x=\frac 1 {D+1}},\\
P_0 &:= P\vert_{F=\frac 1 D} = D^{1-n}\pol({\textstyle \frac 1
{D+1}}).
\end{align}
We will assume $F_1>1$, since the protocol must be meaningful.
The yield after amplifying $\epsilon$ by a factor $t$ is
$\eta^{\log(t)}$, and thus it is justified the introduction of the
coefficient
\begin{equation}\label{eta}
 \eta := \exp \left (
  \frac {\log(P_0)-\log(n)} {\log(F_1)}
 \right ).
\end{equation}
As $F_1<n$ and $P_0\leq1$, then $\eta<e^{-1}$.

We are ready to compare several protocols, which we shall do
progressively increasing the number of discarded pairs.
\begin{itemize}
 \item{$n=2$}. When \eqref{coefsPolinomioVectorial} applies, there
 are just two possibilities. One corresponds to the identity
 permutation an the other is
 \begin{equation}
  \pol(x) = 1+(D-1)x^2.
 \end{equation}
 This corresponds to the original distillation
 protocol discussed in \cite{super}; no surprises. If $D$ is not
 prime there are other possibilities, but $\eta$ is not greater for
 them.
 \item{$n=3$}.
 We must distinguish two cases. If $D$ is odd, the best
 value of $\eta$ is attained with
\begin{equation}\label{polinomio_n_3_mejor}
  \pol(x) = 1+(D^2-1)x^3.
 \end{equation}
 If $D$ is even, however, the best option is
\begin{equation}
  \pol(x) = 1+(D-1)x^2+(D^2-D)x^3.
 \end{equation}
The difference is due to the impossibility of constructing a
suitable $\VM$ in the second case, as we now show. Let $\sset
{\vect u, \vect v}$ be a base of $\VM$. Then, condition \eqref{condicionVM} is
equivalent to
\begin{equation}
  \sum_{i=1}^3
  \begin{vmatrix}
    u_i &u_{3+i} \\
    v_i &v_{3+i} \\
  \end{vmatrix}
  = 0.
 \end{equation}
 In order to obtain \eqref{polinomio_n_3_mejor} the determinants
 appearing in the sum should have an invertible value (see
 \eqref{coefsPolinomioVectorial}), but then they cannot sum up $0$ if D is even.
 \item{$n=4$}.
In this case we have found  that the best polynomial is:
\begin{equation}
  \pol(x) = 1+4(D-1)x^3+(D^3-4D+3)x^4.
\end{equation}
As an example of realizing this case, set $\VM = \lin \sset{\vect u, \vect
v, \vect w}$ with
\begin{align}
\vect u &= (1 , 0 ,0 ,1 ,0 ,1 ,0 ,0) \\
\vect v &= ( 0 , 1 ,0 ,0 ,1, 0, 1 ,0)\\
\vect w &= ( 1 ,1 ,-1,0 ,0 ,0 ,0 ,1)
\end{align}
\end{itemize}

\begin{figure}
\psfrag{F}[Bc][Bc][1][0]{$F$}
\psfrag{F'}[Bc][Bc][1][0]{$F'$}
\psfrag{na}[Bc][Bc][1][0]{$n=2$}
\psfrag{nb}[Bc][Bc][1][0]{$n=3$}
\includegraphics[width=8.5 cm]{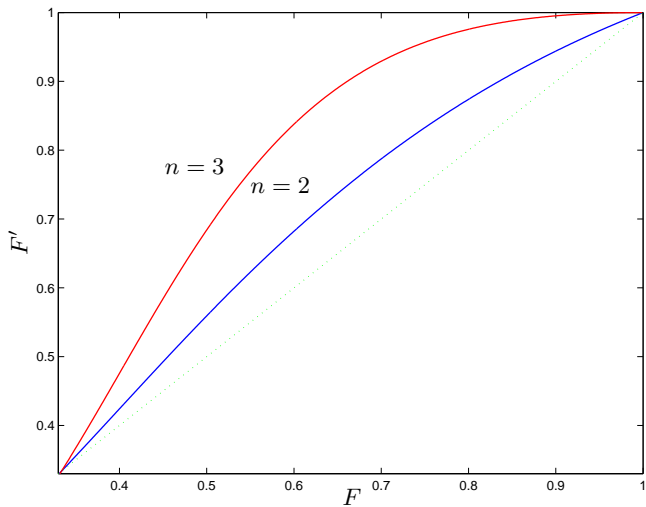}
\caption{
 Evolution of the fidelity for the distillation protocols assisted by
twirling when
 $n=2,3$ for qudits with $D=3$.
} \label{GraphD3}
\end{figure}

  Figure \ref{GraphEta} displays the values of $\eta$ for these
protocols and several dimensions. Note the bad performance of the
case $n=2$ for qubits, which is in fact the most important of all.
On the other hand, qutrits ($D=3$) obtain the best yield among the
proposed protocols, thanks to the advantage of odd dimensionality
(for $n=3$). In connection with this, see also figure \ref{GraphD3}.

One could ask wether it is possible further improvement on $\eta$
by means of increasing $n$. Figure \ref{GraphEta} suggests that
this is not the case, at least for $D>2$ . Exploration shows that
nothing is gained in the case of qubits neither. This result is an
indication of the futility of increasing $n$ with the aim of
improving performance. In \cite{permutations} it is claimed that
protocols with higher $n$ should improve the yield, but apparently
they do not take into account the (strong) reduction in the
probability as $n$ increases.  This idea clarifies figure
\ref{GraphEta}, since from \eqref{probabilidadIdentidad} we expect
$P_0< 2^{n-1}(D+1)^{1-n}$ and then the reduction of the
probability with $n$ is more dramatic as $D$ increases, whereas
the performance gain from $F_1$ is at most linear ($F_1<n$).

\subsubsection{Protocols with $m>1$}
When considering states of higher fidelity, an important advantage
of the proposed protocol for $n=3$ and $D$ odd is that the
derivative of $F'(F)$ vanishes for $F=1$, a qualitative difference
with the $n=2$ case (see figure \ref{GraphD3}). This is important
for states close to the Bell state, since a fidelity $1-\epsilon$
is mapped to $1-O(\epsilon^2)$. We now show how the ratio $\frac n
m = 2$ can be preserved while this desirable characteristic is
added.

\begin{figure}
\psfrag{F}[Bc][Bc][1][0]{$F$}
\psfrag{F'}[Bc][Bc][1][0]{$F'$}
\includegraphics[width=8.5 cm]{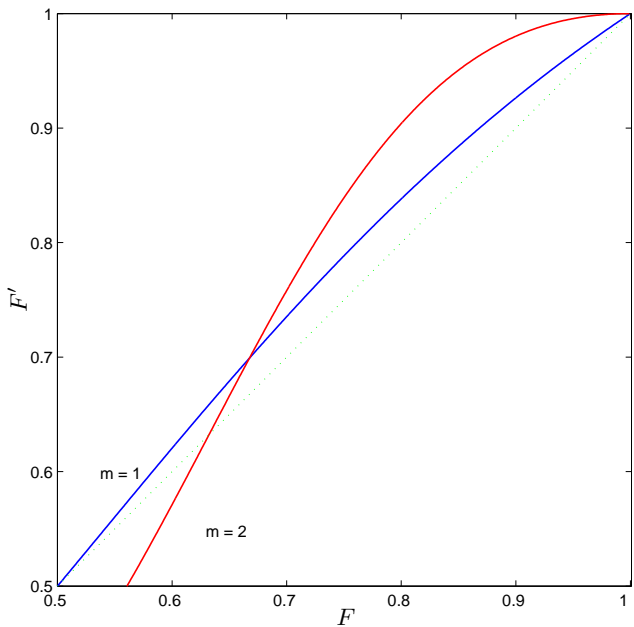}
\caption{
 Evolution of fidelity through the proposed $n=2m$ protocols for
 $D=2$.
} \label{GraphD2}
\end{figure}

\begin{figure}
\includegraphics[width=8.5 cm]{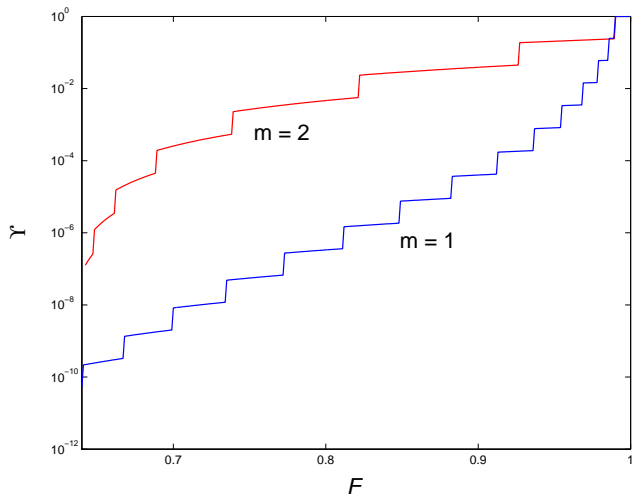}
\caption{
 Yield $\Upsilon$ \eqref{yield} after reaching a fidelity at least $0.99$ through the proposed $n=2m$ protocols
 for $D=2$.
} \label{GraphD2yield}
\end{figure}

Using the definitions in lemma \ref{lemaGrupo}, consider the permutation
\begin{equation}
\pdsum
kl := (\pbsum kl\circ \psum l\circ \psum k )^{-1}\circ \psum
k\circ \pbsum kl.
\end{equation} 
The action of $\pdsum 12$ is
\begin{align}
\perma i j &= \vect i\\
\permb i j &= (j_1+i_2, j_2+i_1,j_3,\dots, j_n); 
\end{align}
We propose for $n=4$ and $m=2$ a protocol in which the permutation
$\ps \M$ of step 2 is 
\begin{equation}
\pbsum 13\circ \pbsum 24 \circ \pdsum
14\circ \pdsum 23.
\end{equation} 
This permutation yields
\begin{equation}
 \pol(x)=1+(D^2-1)x^4.
\end{equation}
The resulting two pairs of qudits will be correlated, thereby providing
us with a good
scenario for hashing, and one could consider an iterative
protocol in which the basic unit were pairs of pairs of qudits
(instead of pairs). We shall keep things simple by choosing $D=2$
and considering the partial traces of each of the pairs (which are
equal due to symmetry of the permutation) in order to obtain the
individual fidelity:
 \begin{equation}
 F' = \frac {F^4} P  \left ( 1 + 4\, c_2(F)^2+4\, c_2(F)^3+7\, c_2(F)^4\right )
 \end{equation}
The results for $D=2$ are displayed in figures \ref{GraphD2} and
\ref{GraphD2yield}.
The yield $\Upsilon$ of figure \ref{GraphD2yield} is calculated step by step
through the following recursion relation:
\begin{equation}
\Upsilon_k = P_k \frac{m}{n} \Upsilon_{k-1}, \ \ \Upsilon_0=1.
\label{yield}.
\end{equation}
Regarding the $n=2$ case, the yield is
greatly increased (four orders of magnitude) even for states quite
near to the fixed point. The drawback is the impossibility of
distillation for $F\lesssim 0.64$, but let us recall that this is 
below the lowest fidelity distillable with a hashing method,
$F\approx 0.81$ \cite{super}. 
The conclusion then is that one
should consider this kind of protocols in the latest steps prior to
hashing.

\subsection{Protocols without twirling}

The use of twirling involves loosing entanglement, and thus the
protocols we have considered so far are  a good starting point
towards more sophisticated ones in which a careful selection of
the permutations avoids the use of twirling techniques and allows
for the distillation of states with fidelity less than $\frac 1 D$, if
$D>2$.

\subsubsection {Quantum Privacy Amplification}

This idea was first explored (for qubits) in \cite{qpa}, where a
quantum privacy amplification scenario was considered. In this
situation the state to be distilled is the average over an ensemble,
not necessarily known, and so the
permutations must work well in general. We shall now generalize
this algorithm to qudits, guided by the main role the vector space
$\VM$ plays, as introduced in the begining of Sect.~\ref{sec5:level1}. 
The proposed generalization is an iteration with $n=2$
and $m=1$ as in the original case. It consists in an alternated
application of two permutations: 
\begin{equation}
\begin{split}
\pbsum 12\circ & \psum 1\circ \psum 2, \\
\pbsum 12\circ & (\psum 1\circ \psum 2)^{-1}. 
\label{permutationsQPA}
\end{split}
\end{equation}

The (relative) simplicity of these operations is a first interesting
point of the algorithm. The choice follows from the intention to
preserve the form of $\VM$ with respect to the known case $D=2$, as
the number of iterations grows. For $s$ iterations, $\M$ in this
case is the $2^{s+1}\times2^{s+1}$ matrix which would follow
 by considering the process as a single iteration,
 with a unique permutation and a unique measurement.

Although the two permutations alternate, it is possible to give a
single recursion relation for every iteration cycle. 
To this end, let us introduce the elements of an alternative Bell basis
as
\begin{equation}
 \bellP{\vect i}{\vect j} := \bell {\vect i}{-\vect j}.
\label{alternative}
\end{equation}
Then, in order to achieve this, it is enough to change the Bell basis to
\eqref{alternative}
after the first
cycle, switch to the original Bell basis after the second, change again
to \eqref{alternative}
after the third one, and so on. 
With this little trick, we get the following recursion relation:
\begin{gather}
  p_{ij}' = \frac 1 P \sum_{k\in\zd} p_{i+k,-i-j-k}\,p_{k,j-k},\\
  P = \frac 1 D \sum_{\tilde k\in\zd} {p_{\tilde k\tilde k}}^2.
 \end{gather}
It is interesting to note that the permutation that can switch
between the two basis is not achievable by local means. If this
were so, we could avoid the use of two different permutations and
still get the same recursion relation, but unfortunately it is not
the case.

\begin{figure}
\includegraphics[width=8.5 cm]{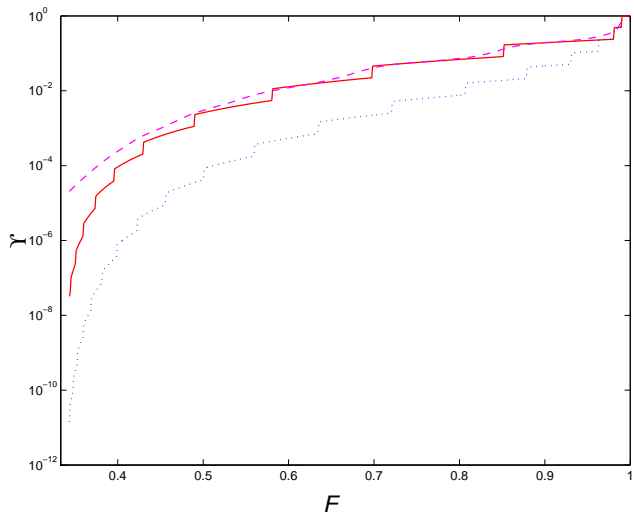}
\caption{
 Yield $\Upsilon$ \eqref{yield} after reaching a fidelity of at least $0.99$ 
through  the protocol proposed in the text: 
for isotropic states (solid line) and as a
 mean over Bell diagonal states (dashed line) compared to the same yield using
 the twirling protocol for $n=2$ (dotted line). The case under study are
 qutrits ($D=3$) and the mean refers to the measure discussed in appendix
 \ref{apendiceMontecarlo} with Monte Carlo. } 
\label{GraphD3Deutsch}
\end{figure}

\begin{figure}
\includegraphics[width=8.5 cm]{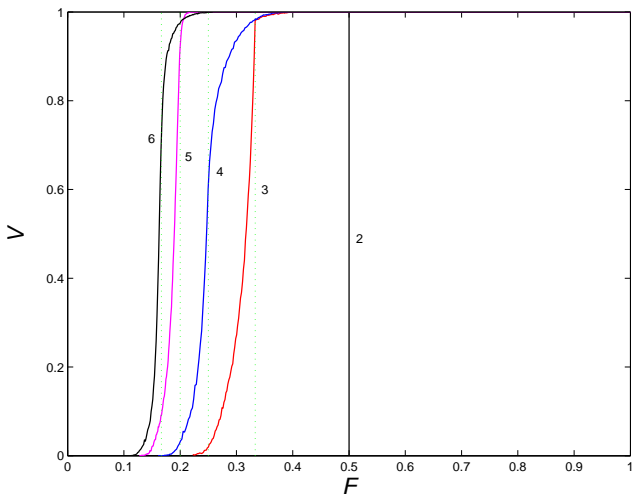}
\caption{
Normalized volume $V$ of distilled Bell diagonal states for the protocol under study, given by permutations \eqref{permutationsQPA} with $D=2,3,4,5,6$.
  The measure is described in appendix \ref{apendiceMontecarlo} 
and uses Monte Carlo.
} \label{GraphDistillable}
\end{figure}

Figure \ref{GraphD3Deutsch} shows the yield of this protocol
compared to the equivalent protocol of the previous subsection 
\ref{sectTwirling}. 
The improvement is clear.
As $D$ increases the results are less
spectacular, however. An important detail is that now the
distillable states are not simply those for which fidelity is
greater than $\frac 1 D$.
As one can check in Fig.~\ref{GraphDistillable}, 
some states over this point are not
distilled whereas other states beneath are. Qubits are the only
ones behaving as expected: the total volume of Bell states with
$F>\half$ is distilled, while those below it are undistilled. 
In any case, the normalized volume of states
showing bad behavior, i.e. those with $F>\frac 1 D$ which are not
distilled, is small if $D$ is prime. This is not so for non prime
$D$'s because, as discussed in section
\ref{sec5:level1}, 
new fixed states
emerge for composite numbers creating undesirable attractors. We find
that these attractors are specially harmful for states near to
heterotropic states.
As a corollary, we show that the permutational approach to distillation is 
more suited to prime $D$'s.

\subsubsection {Distillability}

When the initial state is known, we can make use of this
information to improve the distillation by selecting at each step
the most convenient permutation. Then the question is wether the
protocols we are managing are able to distill any distillable
state. As we lack a working algorithm to decide wether a given state is
distillable, we will compare the normalized volume of distilled states to
that of NPPT states (states with a non positive partial
transpose), since belonging to this set is a necessary condition
for distillability.

We have chosen the following protocol with $n=2$ and $m=1$: At
each step, one of the elements of $\Ps(D,1)$ is applied to both
pairs of qudits before the permutation $p_+^{12}$. The element is
chosen so as to give the best fidelity after the (correct)
measurement.

\begin{figure}
\includegraphics[width=8.5 cm]{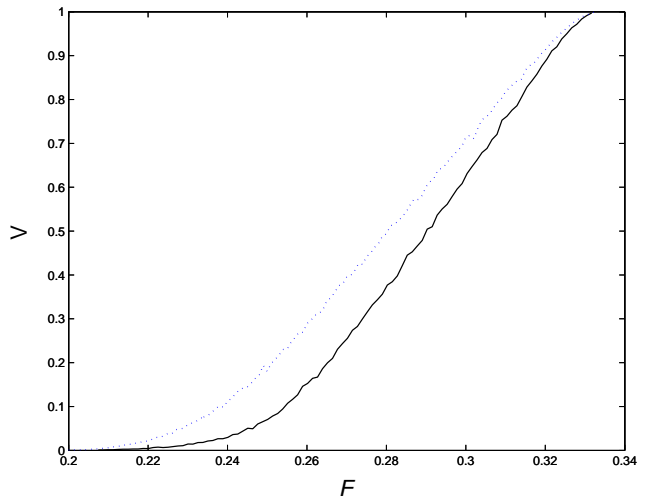}
\caption{
Normalized volume $V$ of distilled Bell diagonal states compared to that of NPPT states 
for $D=3$. The metric and the measuring algorithm are discussed in 
appendix \ref{apendiceMontecarlo} and uses Monte Carlo.} \label{GraphOptimDist}
\end{figure}

Figure \ref{GraphOptimDist} shows the distillation capacities
for $D=3$. 
In general, for $D$ prime the behavior is
good since all states known to be distillable, i.e. those for which
the fidelity is more than $\frac 1 D$,  happen to be distillable
with our protocol. More precisely, 
we have not found computationally any
counterexample of this fact. Not all the NPPT states are
distilled. This is 
perhaps another indication of the existence of
non distillable NPPT states. In the case of composite numbers the
algorithm performs much worse.

\section{Conclusions and Prospects}
\label{sec6:level1}

We have shown that the study of entanglement distillation protocols based on
the recursion method \cite{bennett1} is greatly benefited from the application
of basic number theory concepts when the set $\zdn$ associated to qudits of
arbitrary dimensions $D$    is a module, and not a vector space.
In particular, we have found that a partition of $\zdn$ into divisor classes is
very useful to characterize the invariant properties of mixed Bell
diagonal states under unitary groups that implement local permutations.
This permutations, in turn, are used
in very general distillation protocols based on the recursion method.

We have proposed and study a variety of distillation protocols that fall into
two classes depending on whether we use twirling operations or not at intermediate
steps of the protocols.
When the twirling operations are absent, our distillation protocols amount to
extensions of the quantum privacy amplification protocols \cite{qpa} valid 
for arbitrary qudit dimensions $D$. This is very interesting and relevant for
quantum commmunications with arbitrary large alphabets since they remain secure
and operative even in the presence of quantum noisy channels.

These properties obtained from number theory are not only useful in the analytical
understanding of the protocols, but also facilitate a lot the construction of
numerical methods for their study using Monte Carlo. 
In particular, we have characterized 
how the distillation protocols based on the
recursion method and local permutations are qualitatively and quantitatively
optimal when the dimension of the qudit states $D$ is a prime number.
We leave open the problem of how to construct better distillation protocols 
when $D$ is not a prime number and in this regard, the use of the heterotropic
states  introduced here is a promising tool.

\noindent {\em Acknowledgements}
We acknowledge financial support from a
PFI fellowship of the EJ-GV (H.B.)
and  DGS grant  under contract BFM 2003-05316-C02-01 (M.A.MD.).

\appendix

\section{Properties of the Module $\zdn$}\label{ApendiceVector}

In this appendix we show how many ideas from genuine vector spaces
can be adapted to the module $\zdn$. First of all, we say that an element of
$s\in\zd$ is \emph{invertible} if there exists $s'\in\zd$ such
that $ss'= 1$. If $x\in\n$ is a representant of $s$, this is equivalent to
$\gcd(x,D) = 1$.
 When $D$ is not prime, non invertible elements
other than zero exist (they are multiples of proper divisors of D)
and we need to introduce a work around in the \emph{gaussian
elimination} method, as we shall explain now.

Suppose we are given an element of $\z_D^2$, say $(x,y)$, and we
are asked to get $x=0$ using two elementary transformations:
 \begin{equation}\label{OperacionesElementales}
  (x,y)\overset{\oper 1}{\longrightarrow} (x,x+y),\qquad
  (x,y)\overset{\oper 2}{\longrightarrow} (x+y,y),
 \end{equation}
The algorithm turns out to be quite simple. Consider for a moment the
arbitrary ordering in $\zd$ $0<1<\cdots<{D-1}$. At each step, if
$x\leq y$ use $\oper 1$ to get $0 \leq y<x$, proceed inversely on
the contrary. Clearly, $x=0$ or $y=0$ is reached in a finite
number of steps. If $y=0$, just apply $\oper 1$ once and $\oper 2$
$D-1$ times.

We shall use gaussian elimination, with the aid of the above
trick, to convert any matrix $M$ of size $p\times q$ into a very
simple one. Suppose $p\leq q$; the converse case is similar.
Summing one row to another amounts to take the product (from the
left) with a $p\times p$ invertible matrix. The same is true for
columns (from the right, $q\times q$). Using this elementary
operations we can obtain:
\begin{equation}\label{descompMatriz}
 C = AMB, \qquad C=
 \begin{bmatrix}
  D & 0
 \end{bmatrix},
\end{equation}
where $A$ and $B$ are invertible and $D$ is a diagonal matrix.

With this tool at hand, we are ready to start with our analysis. We
adopt the usual definition of linear independence for a finite
subset of $\zdn$, but the following is more surprising:
\begin{defn}\label{defRank}
 Consider any $M\in \Mat_{p\times q}(\zd)$ and let $S_r$ be the set of
 all $r\times r$ minors of $M$, $r\in R:=\sset{1,\dots, \min(p,q)}$.
 The rank of M is defined as
 \begin{equation}
  \rango(M) := \max \; \sset{0} \cup \set {r\in R}{\mcd(S_r)=1}.
 \end{equation}
\end{defn}
The rank of a matrix does not vary when we apply the elementary
operations discussed above (see \eqref{defmcdAlternativa} and take
into account that $d|x$ and $d|y$ iff $d|x$ and $d|x+y$). As
expected, a square matrix is invertible iff its rank is maximal.
The following statement clarifies this strange definition:
\begin{prop}\label{lemaLIvsRank}
 The rows of a matrix $M\in \Mat_{p\times n}(\zd)$ form a l.i. set iff $\rango(M)=p$.
\end{prop}
\begin{proof}
 Recall decomposition \eqref{descompMatriz} for $M$ (we will use directly the notation there)
 and consider any $\vect v\in\zd^p$:
 \begin{equation*}
  \vect v\transp M = 0 \iff
  \vect v\transp A^{-1}CB^{-1} = 0 \iff
  {\vect v'}\transp C = 0,
 \end{equation*}
 where $\vect v = A\transp \vect v'$.
 This shows that the rows of $M$ form a l.i. set iff the rows of
 $C$ do. On the other hand, $\rango(C)=\rango(M)$, and for the
 matrix $C$ the statement is trivial.
\end{proof}

Given a set $S=\sset{\vect v_1,\dots,\vect v_k}\subset\zdn$, we
will denote by $\lin S$ the subset of $\zdn$ spanned by the
elements of $S$, that is, the set of linear combinations of the
vectors in $S$. Clearly, if $S$ is l.i. $\sharp \lin S=D^k$, and
so $k\leq n$. Not surprisingly, any l.i. set which spans $\zd$
will be called a base of $\zdn$. The usual definition of subspace
does not work, however, and we introduce in its place the
following:
\begin{defn}\label{defSubesp}
 A set $V\subset\zdn$ is said to be a subspace of $\zdn$ if
 $V=\sset{0}$ or if there exists a set $S = \sset{\vect v_1,\dots,\vect v_k} \subset\zdn$
 such that it is l.i. and $\lin S=V$. Such a set is called a base of $V$,
 and its cardinality is the dimension of $V$ $(\dim V)$.
\end{defn}
Dimension is well defined since $\sharp V=D^{\sharp S}$ forbids
the possibility of two bases of different cardinality.

\begin{prop}\label{lemaExtensibilidad}
 Given a set $S=\sset{\vect v_1,\dots,\vect v_k}\subset\zdn$
 which is linearly independent, there exists a set $S'=\sset{\vect v_{k+1},\dots,\vect
 v_n}$ such that $S\cup S'$ is a base of $\zd$.
\end{prop}
\begin{proof}
 Let $M$ be a $k\times n$ matrix such that its rows are the
 elements of $S$. We recall \eqref{descompMatriz} but rewrite it in
 terms of $n\times n$ square matrices:
 \begin{equation}
 \begin{bmatrix}
  D & 0 \\ 0 & 0
 \end{bmatrix}
 =
 \begin{bmatrix}
  A & 0 \\ 0 & 1
 \end{bmatrix}
\begin{bmatrix}
  M \\ 0
 \end{bmatrix}
 B
 \end{equation}
 Now consider the following:
 \begin{equation}
\begin{bmatrix}
  M \\ M'
 \end{bmatrix}
 =
 \begin{bmatrix}
  A^{-1} & 0 \\ 0 & 1
 \end{bmatrix}
 \begin{bmatrix}
  D & 0 \\ 0 & 1
 \end{bmatrix}
 B^{-1}
 \end{equation}
 It is enough to construct $S'$ with the rows of $M'$.
\end{proof}
\begin{cor}\label{lemaExtensibilidadSubesp}
 Given a subspace $V$ of dimension $d$ and a set $S=\sset{\vect v_1,\dots,\vect v_k}\subset V$
 which is linearly independent, there exists a set $S'=\sset{\vect v_{k+1},\dots,\vect
 v_d}$ such that $S\cup S'$ is a base of $V$.
\end{cor}
\begin{proof}
 Select any base of $V$ and consider the components of vectors
 respect to that base as elements of $\zd^d$.
\end{proof}

We adopt the usual definition and notations for the scalar product
and orthogonality.
\begin{prop}
 The subset of $\zdn$ orthogonal to a subspace $V$ (that is,
 $V^\bot$) is a subspace. Moreover, $\dim V + \dim V^\bot = n$.
\end{prop}
\begin{proof}
 Let $d = \dim V$ and let $M$ be a $d\times n$ matrix such that its rows form a base of $V$.
 We use decomposition \eqref{descompMatriz} (again). For any $\vect v\in\zdn$
 \begin{equation}
  M \vect v = 0 \iff
   A^{-1}CB^{-1}\vect v = 0 \iff
   C (\vect v') = 0,
 \end{equation}
 with $\vect v = B \vect v'$. The set of the $\vect v'$-s
 verifying the equation is clearly a subspace of the expected
 dimension.
\end{proof}

\section{Cardinality of $\particion d D n$: generalized Euler's
totient function}\label{ApendicePhi}

This appendix is devoted to the proof of lemma \ref{lemaPhi}. We
start with  part 2, which can be rewritten as:
\begin{equation}
 \sharp \particion 1 D n = \sharp \particion d {dD} n \qquad
 \forall\: D\geq2,n\geq 1, d\geq 1
\end{equation}
This is equivalent to the existence of a one to one mapping from
$\particion 1 D n$ onto $\particion d {dD} n$. Consider the
mapping:
\begin{align}
\mu: \z^n&\longrightarrow\z^n \\
\vect v&\longrightarrow d\vect v
\end{align}
$\mu$ induces a mapping $\overline{\mu}: \zdn\rightarrow
\z_{dD}^n$, which is well defined and one-to-one because $x=y
\pmod D \Leftrightarrow dx=dy \pmod {dD}$. 
From \ref{defmcdAlternativa} we learn that $\forall\:\vect
v\in\z^n$
\begin{equation*}
  d|\mcd(\overline{\mu}(\overline{\vect v})),
\end{equation*}
where $\overline{\vect v}$ is the result of mapping $\vect v$ in
$\zdn$. Since for any $x\in\z$ and $d'\in\divis D$ we have
$d'|x\Leftrightarrow d'd|dx$, it follows that
\begin{equation}
 d\mcd(\overline{\vect v}) = \mcd(\overline{\mu}(\overline{\vect
 v})),
\end{equation}
which implies that $\particion 1 D n$ is mapped into $\particion d
{dD} n$. Since for any element of $\particion d {dD} n$ there exists a
suitable $\vect v$, the mapping is onto.

Now proving  part 1 of the lemma is easy. Start with
 \begin{equation}
\begin{split}
 \indep n D &= D^n - \sum_{d\in \divis D-\sset 1}
   \sharp \particion d D n \\
    &= D^n - \sum_{d\in \divis D-\sset 1}
   \indep n {\frac D d}.
\end{split}
 \end{equation}
Changing the index, we get a beautiful recursive relation:
 \begin{equation}
  \indep n D = D^n - \sum_{d\in \divis D-\sset D} \indep n d.
 \end{equation}
With some  algebra on this expression it is possible to show that $\indep n D$
is a multiplicative function: A function $f:\n\rightarrow\n$ is said to be 
multiplicative if
$f(nm)=f(n)f(m)$ $\forall\:n,m\in\n$ such that $\gcd(m,n)=1$.
Thus, we only have to solve the recursion for $D = p^q$, $p$ prime, but this
poses no difficulty:
\begin{equation}
 \indep n {p^q} = p^{nq}-p^{(n-1)q},
\end{equation}
from which \eqref{valorphi} follows.

Part 3 of the lemma 
is nothing but the recursion relation just constructed.

\section{Generators of $\Ps$}\label{ApendiceGeneraPs}

In order to study $\Ps$ it is preferable to consider its elements as
permutations over
$\zdn\times\zdn$, and
so we change the notation
\begin{align}
\vect x &\longrightarrow \perm(\vect x), \quad &\vect x \in\zddn
    \intertext{for}
(\vect i,\vect j) &\longrightarrow (\perma i j,\permb i j), \quad
&\vect i,\vect j \in\zdn.
\end{align}
where the correspondence is the same as in
\eqref{CorrespondenciaZddnZdn}.
\begin{lem}\label{lemaGrupo}
 $\Ps$ is generated by its following elements:
 \begin{align*}
  \psum 1, \quad\text{ with }\quad
   \perma i j &= \vect i,
   \\ \permb i j &= (i_1 + j_1, j_2, \dots, j_n);\\
  \pex 1, \quad\text{ with }\quad
   \perma i j&= (j_1, i_2, \dots, i_n),
   \\ \permb i j&= (-i_1, j_2, \dots, j_n);\\
  \pbsum 1 2, \quad\text{ with }\quad
   \perma i j &= (i_1+i_2, i_2,\dots, i_n),\\
  \permb i j &= (j_1, j_2-j_1,\dots, j_n); \\
  \pswap l m, \quad\text{ with }\quad
   \perma i j &= (\dots, i_{l-1}, i_m, i_{l+1}, \dots, \\ 
              & i_{m-1}, i_l, i_{m+1}, \dots ), \\
  \permb i j &= (\dots, j_{l-1}, j_m, j_{l+1}, \dots, \\
             & j_{m-1}, j_l,
   j_{m+1},\dots), \ \  l,m = 1, \dots, n.
 \end{align*}
\end{lem}

\begin{proof}
In order to proof this, first let us define
 \begin{align}
  \psum i &:= \pswap 1 i \circ \psum 1 \circ \pswap 1 i,\\
  \pex i &:= \pswap 1 i \circ \pex 1 \circ \pswap 1 i,\\
  \pbsum i j &:= \pswap 1 i \circ \pswap 2 j \circ
   \pbsum 1 2 \circ \pswap 1 i \circ \pswap 2 j,
 \end{align}
 with 
$i,j = 1,\dots, n$. Consider any $p\in\Ps$; our goal is to act
 from the left and from the right with these permutations
 till we get the identity, which is equivalent to the
 statement of the lemma. We shall use the matrix representation of
 the permutations, and the process is a suitable gaussian
 elimination similar to the one used in appendix
 \ref{ApendiceVector}. The difference is that now we cannot
 perform freely any sum of lines or columns, but only those
 which have associated a permutation in the above set.

 To work around this problem, in place of \eqref{OperacionesElementales}
  we consider
 \begin{equation}
  (x,y)\overset{\oper 1}{\longrightarrow} (x,x+y),\qquad
  (x,y)\overset{\oper 3(e)}{\longrightarrow} (ey,x),\quad e=\pm 1
 \end{equation}
 Since $\oper 2$ can be constructed suitably combining $\oper 1$
 and $\oper 3(-1)$, only the case $e=1$ is really different, but
 adapting the algorithm is straightforward.
 The point is that $\psum i$ and $\pbsum i j$ can be attached to
 $\oper 1$, $\pex i$ to $\oper 3(-1)$ and $\pswap i j$ to $\oper 3(1)$, 
 with care in the case of $\pbsum i j$ and $\pswap i j$ for their
 additional  effects.

 To perform the elimination in an element of $\Ps$ with associated matrix $\M$,
  start working over the first
 column (permutations act thereby from the left). Using $p_+^i$ and $p_{ex}^i$
 make zero the elements $\M_{n+1,1}$ to $\M_{2n,1}$, and then
 use $p_+^{1i}$ and $p_{swap}^{1i}$ till just $M_{11}$ is nonzero
 in the first column. The process must be repeated for the first
 row (this time permutations act from the right). Now let us deal with the
 second column, first making zero the elements $M_{n+2,2}$ to
 $M_{2n,2}$, afterwards the elements $M_{3,2}$ to $M_{n,2}$. Do
 the same for the second row. The process must be carried out for the first $n$ rows and
 columns, till we get something of the form
 \begin{equation}\label{}
  \begin{bmatrix}
   D & T_1 \\
   T_2 & M
  \end{bmatrix},
 \end{equation}
  with $D$ diagonal, $T_1$ lower triangular and $T_2$ upper triangular.
  But in fact applying the condition \eqref{CondicionGrupoM} forces
 $T_1 = T_2 = 0$ and $M=D^{-1}$. Now we note that
\begin{equation}\label{}
  \begin{bmatrix}
   D & 0 \\
   0 & D^{-1}
  \end{bmatrix} =
  \begin{bmatrix}
   1 & 0 \\
   -D & 1
  \end{bmatrix}
  \begin{bmatrix}
   1 & 0 \\
   D-1 & 1
  \end{bmatrix}
  \begin{bmatrix}
   1 & 1 \\
   0 & 1
  \end{bmatrix}.
 \end{equation}
 Since the matrices in the right side are trivially constructed with
 the given set of generators, this ends the proof.
\end{proof}

\section{Proof of theorem \ref{thmPlocal}}\label{ApendiceProofThm}

(This proof uses the notation and results of appendix
\ref{ApendiceGeneraPs})

(1) Since $\Pt$ is invariant in the product group trivially, we
prove both sides of the inclusion, starting with
$\Pl\subset\Pt\ltimes\Ps$.

Later we define local unitary operators implementing $\Pt$ (see
\eqref{ComoTraslacion}), and so
 we just bother about those $U\in\Ubl(D,n)$ that leave $\bellketbra 0
 0$ invariant. Moreover, as the global phase is unimportant, we
 select for the analysis those operators for which $U\bell 0 0 = \bell 0 0$.
But $U$ is local, and then
these constraints are equivalent to
$U=U_{\rm A}\otimes U_{\rm A}^\ast$ (conjugation respect to the computational
basis). We can thus write:
\begin{equation}
 U\biket{\vect i}{\vect j} = \suma {\vect k\vect l}
 A_{\vect k\vect i}A_{\vect l\vect j}^* \,\biket {\vect k}{\vect l},
 \qquad \suma {\vect k} A_{\vect i\vect k} A^\ast_{\vect j\vect
 k} = \delta(\vect i- \vect j)
\end{equation}
 with $A$ a unitary matrix on a single party (Alice or Bob). Using \eqref{Bell}, we get
 \begin{equation}
  U\bell{\vect i}{\vect j} = \suma {\vect k\vect l\vect m\vect n}
  \fase{\vect k \cdot \vect i - \vect m \cdot \vect l}
  A_{\vect l\vect k}A_{\vect l - \vect n,  \vect k - \vect j}^\ast
  \,\bell {\vect m}{\vect n}.
 \end{equation}
 On the other hand, the action of $U$ involves a permutation over
 Bell states:
 \begin{equation}
  U\bell{\vect i}{\vect j} = \fasePerm i j\,\bell {\perma i j}{\permb i j}.
 \end{equation}
 Identifying both expressions (Bell states are orthogonal):
 \begin{multline}
  \fasePerm i j
  \,\delta (\vect m - \perma i j)
  \,\delta (\vect n - \permb i j)
  = \\
  \suma {\vect k\vect l}
  \fase{\vect k \cdot \vect i - \vect l \cdot \vect m}
  A_{\vect l\vect k}A_{\vect l - \vect n,  \vect k - \vect j}^\ast.
  \end{multline}
 Act on both sides of this equation with the Fourier operator 
$\sumatext {\vect m\vect n} \fase{\vect r \cdot \vect m}A_{\vect
r-\vect n,\vect s-\vect
 j}$ to obtain:
 \begin{equation}\label{recurrencia}
  A_{\vect r - \permb i j,  \vect s - \vect j}
  =
  \fase{\vect s \cdot \vect i - \vect r \cdot \perma i j}
  \,\fasePerm i j^\ast A_{\vect r, \vect s}.
 \end{equation}
 Choose any $\vect r, \vect s$ such that $A_{\vect r\vect s}\neq
 0$, interpret this equation as a recurrence relation and consider the commutative diagram:
 \[
 \begin{array}{ccc}
  A_{\vect r\vect s}
  & \longrightarrow &
  A_{\vect r - \vect \pi (\vect i, \vect j),  \vect s - \vect j}
  \\
  \Big\downarrow
  & &
  \Big\downarrow
  \\
  A_{\vect r - \vect \pi (\vect i', \vect j'),  \vect s - \vect j'}
  & \longrightarrow &
  A_{\vect r - \vect \pi (\vect i, \vect j)-\vect \pi (\vect i', \vect j'),
    \vect s - \vect j- \vect j'}
 \end{array}
 \]
 Switching again to the $\zddn$ notation, the commutation condition is
 \begin{equation}\label{conmutacion}
   \vect x\prodSym\vect x' = \perm(\vect x)\prodSym\perm(\vect x')
 \end{equation}
 for any $\vect x, \vect x' \in \zddn$. Thus, $\perm\in\Ps$.

 We now show that $\Pt\ltimes\Ps\subset\Pl$. Thanks to lemma \ref{lemaGrupo},
 it is enough to construct a few permutations by means of unitary local operators.
 We start with translations. Choosing
 \begin{align}
  U_{\rm A}\,\ket{\vect i} = \fase{\vect i \cdot \vect a}\,\ket{\vect
  i},&\qquad
  U_{\rm B}\,\ket{\vect i} =
   \ket{\vect i - \vect b},\label{ComoTraslacion}
          \intertext{with $\vect a, \vect b \in \zdn$, the effect is}
   \perma i j = \vect i + \vect a,&\qquad \permb i j = \vect j + \vect
   b.\notag
           \intertext{This is not the only subgroup easily generated, we have also
 that}
  U_{\rm A}\,\ket{\vect i} = \ket{S \vect i}, &\qquad
  U_{\rm B}\,\ket{\vect i} = \ket{S \vect i},
           \intertext{with $S \in \zdn\times\zdn$ and invertible give}
  \perma i j = (S^t)^{-1} \vect i,&\qquad \permb i j = S \vect j.\notag
 \end{align}
Both of these results can be checked with a few manipulations in
 \eqref{nBell}. Since $\pswap l m$ is physically trivial and $\pbsum l m$ is contained
  in the last construction, the only permutations
 we have not still covered are $\psum 1$ and $\pex 1$, but as these involve only
 the first pair of qudits we can fix $n=1$ in \eqref{recurrencia} and try the
 \emph{ansatz} $A_{lk} = \fase{\eta(i, j)}$, with $\eta(i,j)\in\mathbf{R}/D$.
 This results in:
 \begin{equation}
  \eta(l - \permas i j,  k - j) =
  \eta (l,k) + ki - l\permbs i j - \barfasePerms i j,
 \end{equation}
 where $\fase{\barfasePerms i j} := \fasePerms i j$. Solutions to this equation require
 $\eta$ to be a second order polynomial, limiting the permutation to:
 \begin{equation}\label{x}
  \permas i j = aj+b\permbs i j, \quad  \permbs i j = -a^{-1}(i+cj),
 \end{equation}
 where $a, b, c \in \zd$, $a$ invertible. A compatible choice for
 $\eta$ is
 \begin{align}
  \eta(i,j) &= aij+\frac{b}{2}i^2+\frac{c}{2}j^2.
 \end{align}
The permutations we were searching for belong to the set of \eqref{x}.

(2) In order to proof the second part of the theorem it is enough
to analyze which are the realizations of the identity permutation.
Going back to \eqref{recurrencia} and fixing $\perma i j = \vect i$
and $\permb i j = \vect j$ we find the equation
\begin{equation}
 A_{\vect r - \vect j,  \vect s - \vect j}
  =
  \fase{\vect s \cdot \vect i - \vect r \cdot \vect i}
  \,\fasePerm i j^\ast A_{\vect r, \vect s}.
 \end{equation}
Modulo a global phase \eqref{globalphase}, the solutions are exactly of the form:
\begin{align}
 \phi(\vect i,\vect j) &= \fase{\vect a\cdot \vect i + \vect
 b\cdot \vect j} \\
  A_{\vect r \vect s} &= \fase{\vect b\cdot \vect s}
  \delta(\vect s - \vect r - \vect a)
\end{align}
where $\vect a, \vect b \in \zdn$. But this is $\ubi x$ in \eqref{globalphase} with
\begin{equation}
 \vect x = (b_1,\dots,b_n,-a_1,\dots,-a_n).
\end{equation}

\section{Order of $\Ps$}\label{ApendiceCardinalPs}
 In this appendix we offer a proof of theorem
\ref{thmCardinalPs}.

We first note that, except for a sign, $\vect u_i$ and $\vect v_i$
play interchangeable roles. Thus it is enough to consider a case
with $t=0$ (if $t\geq 1$, suitable exchanges between $\vect u$-s
and $\vect v$-s and sign adjustments will be enough). We shall
consider two cases separately, depending on wether $r<n$. In both
cases the target is to find out in how many ways a new vector can
be included in the set. Such a vector must fulfill
\eqref{Canonicas} and be linearly independent respect to the
initial set.

Suppose $r<n$. We would like to know how many vectors can take the
role of $\vect u_{r+1}$. Let $S=\sset{\vect u_1,\dots,\vect
u_r,\vect v_1,\dots,\vect v_s}$ and $V = \lin \set {\sym \vect
v}{\vect v \in S}$. From \eqref{Canonicas} we have $\vect u_{r+1}\in
V^\bot$ (and no further conditions), so let $S'=\sset{\vect
u_{s+1},\dots,\vect u_r, \vect w_1, \dots, \vect w_{2(n-r)}}$ be a
base of $V^\bot$. We claim that $S\cup S'$ is l.i., that is, the
equation
\begin{equation*}
 \sum_{i=1}^{r-s} a_i \vect u_{s+j}+\sum_{i=1}^{2(n-r)} b_i\vect w_i + \sum_{i=1}^s (c_i\vect u_i+d_i\vect v_i) = 0
\end{equation*}
holds only if all the scalars are zero. This is so because taking
the scalar product with $\sym\vect u_k$ ($k\leq s$) we get
$d_k=0$; using $\sym\vect v_k$, $c_k=0$. The rest of the scalars
must be zero because $S'$ is l.i.  Therefore there are
$D^{r-s}\indep {2(n-r)} D$ suitable vectors, since we can choose
any combination of the form
\begin{equation*}
 \vect u_{r+1} = \sum_{i=1}^{r-s} a_i \vect u_{s+j}+\sum_{i=1}^{2(n-r)} b_i\vect w_i
\end{equation*}
for which $\mcd(\sset{b_1,\dots,b_{2(n-r)}})=1$ (this is why the
factor $\indep {2(n-r)} D$ appears, see \eqref{defPhi}).

Now suppose $r=n$, $s<n$. We pursue $\vect v_{s+1}$. Let
$S=\sset{\vect u_1,\dots,\vect u_{s},\vect u_{s+2},\dots,\vect
u_n,\vect v_1,\dots,\vect v_s}$ and $V = \lin \set {\sym \vect
v}{\vect v \in S}$. From \eqref{Canonicas} we have $\vect
v_{s+1}\in V^\bot$ and $\vect u_{s+1}\prodSym\vect v_{s+1} = 1$.
Let $S'=\sset{\vect u_{s+1},\dots,\vect u_n, \vect w}$ be a base
of $V^\bot$ and let $s := \vect u_{s+1}\prodSym\vect w$. We first
show that $s$ is invertible. Let $V'=\lin \sset{\sym\vect
u_1,\dots,\sym\vect u_n, \sym\vect v_1,\dots,\sym\vect v_s}$. If
$s$ is non invertible, choose any $k\neq 0$ such that $ks=0$. Then
$k\vect u_{s+1}\prodSym\vect w=0$ implies $k\vect w\in
V'^\bot=\lin\sset{\vect u_{s+1},\dots,\vect u_n}$, but this is not
possible because $S'$ is linearly independent. We now show that
$S\cup\sset{\vect u_{s+1},\vect w}$ is l.i. If it is not, then
$\vect w \in \lin (S\cup\sset{\vect u_{s+1}})$, but this in
turn implies $\vect u_{s+1}\prodSym\vect w=0$, which again is false.
Therefore, there are $D^{n-s}$ suitable vectors, since we can
choose any of the following combinations:
\begin{equation*}
 \vect v_{s+1} = s^{-1}\vect w +  \sum_{i=1}^{n-s} c_i \vect
 u_{s+i}.
\end{equation*}

With this, part 1 of the theorem is proved. For part 2, it
only remains to count. There are $\indep {2n} D$ possible values
for $\vect u_1$. If $\vect u_1$ is fixed, there are $D\indep {2n}
D$ possible elections for $\vect u_2$. Continuing this way, one
gets the desired result.

\section{Recursion Relations for Distillation Protocols}\label{ApendiceDestilacion}

In this appendix we derive an expression for the final state of
the remaining pairs of qudits when the procedure of section
\ref{sec5:level1} has been successfully
performed. We will use the same notation found there.

So let us define for $\vect x, \vect y \in \zddn$
\begin{align}\label{}
 \multirho n_{\vect x \vect y} &:= \sbellbra {\vect x} \multirho n \sbell {\vect
 y},
\end{align}
from which $
 p_{\vect x} = \multirho n_{\vect x \vect x}$. After the permutation with associated matrix $M$ and
phase function $\phi$ the state is
\begin{equation}\label{defMultirhoIntermedio}
 {\multirho n}':=\sum_{\vect x, \vect y \in \zddn} \phi(\vect x)\phi^\ast(\vect y) \,\multirho n_{\vect x\vect y}\, \sbellketbradif {M\vect x} {M\vect y}.
\end{equation}
The measurement is performed in the computational basis (for the
last $n-m$ pairs), and the rest of pairs are kept only if this
measurement coincides for each of the measured pairs (if Alice
measures $\ket 3$, so does Bob for the corresponding qudit). Going
back to \eqref{Bell}, this means that $j$ is zero for each of the
pairs. Therefore after the measurement and taking the partial
trace over the measured pairs the state of the first $m$ pairs is:
\begin{equation}\label{estadofinalBasico}
\multirho m = \frac 1 P \sum_{\vect k\in\zd^{n-m}}
\bellbra {\vect k} {\vect 0} \,{\multirho n}'\,
\bell {\vect k} {\vect 0},
 \end{equation}
where the Bell states must be understood to belong to the space of
the last $n-m$ pairs and $P$ is the probability of having obtained
the suitable measurement. Calculating it amounts to take the total
trace:
\begin{equation*}\label{ProbabilidadBasica}
P = \sum_{\vect x\in\zddm}\sum_{\vect k\in\zd^{n-m}} \left
(\sbellbra {\vect x}\otimes\bellbra {\vect k} {\vect
0}\right)\,{\multirho n}'\, \left (\sbell {\vect x}\otimes\bell
{\vect k} {\vect 0}\right).
 \end{equation*}
Inserting definition \eqref{defMultirhoIntermedio}, we get
\eqref{probabilidadResultado}.

The state of the system before the measurement can also be
expressed
\begin{equation*}\label{}
 {\multirho n}'=\sum_{\vect x,\vect y\in \zddn} \phi(\M^{-1}\vect x)\phi^\ast(\M^{-1}\vect y)
 \,\multirho n_{\M^{-1}\vect x,\M^{-1}\vect y}\,
 \sbellketbradif {\vect x}{\vect y}.
\end{equation*}
Inserting this expression in \eqref{estadofinalBasico}:
\begin{multline}\label{estadoFinalIntermedio}
 \sbellbra {\vect x} \multirho m \sbell {\vect y} = \frac 1 P
  \sum_{\vect k \in \zd^{n-m}} \phi(\M^{-1}(\vect {\hat k}+\vect {\bar x})) \\ \phi^\ast(\M^{-1}(\vect {\hat k}+\vect {\bar y}))
    \,\multirho n_{\M^{-1}(\vect {\hat k}+\vect {\bar x}),\M^{-1}(\vect {\hat k}+\vect {\bar y})}
\end{multline}
where $\vect x,\vect y\in \zddm$, $\vect {\bar{x}}$ and $\vect
{\bar{y}}$ are defined as in \eqref{defxBarrada} and:
\begin{equation*}
\vect {\hat k} :=
(\underset{m}{\underbrace{0,\dots,0}},k_1,\dots,k_{n-m},\underset{n}{\underbrace{0,\dots,0}}).
\end{equation*}
 With the definition for $\VM$
given in section \ref{sec5:level1}, we have
\begin{multline*}
\sbellbra {\vect x} \multirho m \sbell {\vect y} =  \frac 1 P
  \sum_{\vect z \in \VM} \phi(\sym\vect z+\M^{-1}\vect {\bar x})\\
\phi^\ast(\sym\vect z+\M^{-1}\vect {\bar y})
    \,\multirho n_{\sym\vect z+\M^{-1}\vect {\bar x},\sym\vect z+\M^{-1}\vect {\bar
    y}}.
\end{multline*}
\eqref{recurrenciaProbabilidades} follows setting $\vect x = \vect
y$.

\section{Monte Carlo measuring}\label{apendiceMontecarlo}

We introduce a suitable metric in the space of Bell diagonal
states in order to perform several measures. For simplicity, we
have chosen the metric induced by mapping physical states into
Euclidean space taking the eigenvalues as coordinates.

We have chosen a Monte Carlo approach to perform the measurements.
This approach consists in randomly generating points of the space according
to the measure on that space, 
and counting how many of them are
inside the measured set.

In our case, numerically implementing such a
measure is not difficult if fidelity is not low. Consider a Bell
diagonal state of fidelity $F$. There are $D^2-1$ free coordinates
(eigenvalues) $\lambda_i$ subject to the constraints
\begin{equation}
 0 \leq \lambda_i \leq F,\qquad \sum_i \lambda_i = 1-F.
\end{equation}
The random generation is achieved as follows.
We take for
each point $D^2-2$ real random variables $x_i$ uniformly
distributed in $[0,1]$ $(i=1,\dots, D^2-2)$. Defining $x_0:=0$ and
$x_{D^2-1}:=1$, we set $\lambda_i = (1-F) (x_{i+1}-x_i)$. If
$\lambda_i>F$ for any $i$, we simply discard the point. Else, it
belongs to the space of interest. Then it is checked wether it belongs
to the measured set, by running the appropiate algorithm. For example, if we are checking
distillability through a given protocol, this is the moment were the protocol is numerically
simulated till the point converges.


\end{document}